\documentclass[11pt]{article}
\usepackage{amsmath}
\usepackage{latexsym}
\usepackage{amssymb}
\usepackage{epsfig}
\usepackage{xspace}
\usepackage{mdwlist}

\newtheorem{theorem}{Theorem}[section]

\newtheorem{Lem}[theorem]{Lemma}
\newtheorem{claim}[theorem]{Claim}

\newtheorem{Def}[theorem]{Definition}

\newenvironment{myproof}{\noindent {\sc Proof:}}{$\Box$}

\newcommand\pr{\mbox{\bf Pr}}
\newcommand\av{\mathop{\mathbb{E}}}
\newcommand\E{{\mathbb{E}}}
\newcommand\var{\mbox{\bf Var}}

\newcommand \poly{\mbox{poly}}

\def\epsilon{\varepsilon}
\def\lap{{\mathcal L}}
\newcommand\qi{q_i}
\newcommand\qhi{\widehat{y}_i}
\newcommand\qbi{\bar{y}_i}
\newcommand\qti{\widetilde{y_i}}
\newcommand\halfqi{y_i}

\newcommand\round{f}
\newcommand\dense{\qhi}

\newcommand\good{\textrm{Good}}
\newcommand\cross{\textrm{Cross}}
\newcommand\inc{\textrm{Inc}}
\newcommand\cut{\textrm{Cut}}

\newcommand\Recursive{{\`sc Recursive-walk}\xspace}
\newcommand\Balance{{\sc Balance}\xspace}
\newcommand\Threshold{\textsc{Threshold}\xspace}
\newcommand\Find{\textsc{Find-threshold}\xspace}
\newcommand\Simple{{\sc Simple}\xspace}
\newcommand\CutOrBound{{\sc CutOrBound}\xspace}
\newcommand\otilde{\widetilde{O}}

\newcommand\soto{f\xspace}
\newcommand\volls{\omega}

\def\maxcut{\textsc{MaxCut}\xspace}
\def\Dhalf{D^{1/2}}
\def\Dhalfinv{D^{-1/2}}
\def\vol{\text{Vol}}

\begin{document}
\begin{titlepage}
\title{Combinatorial Approximation Algorithms for {\sc MaxCut} using Random Walks}
\date{}
\author{
Satyen Kale\\
Yahoo! Research\\
4301 Great America Parkway, \\
Santa Clara, CA 95054\\
{\tt skale@yahoo-inc.com} \and
C. Seshadhri \\
IBM Almaden Research Center\\
650 Harry Road, \\
San Jose, CA 95120\\
{\tt csesha@us.ibm.com}}
\maketitle

\thispagestyle{empty}
\begin{abstract}

We give the first combinatorial approximation algorithm for \maxcut that beats
the trivial $0.5$ factor by a constant. The main partitioning procedure is very intuitive, natural, and easily described. It essentially performs a number of random walks and aggregates the information to provide the partition. We can control the running time to get an approximation factor-running time tradeoff. We show that for any constant $b > 1.5$, there is an $\otilde(n^{b})$ algorithm that outputs a $(0.5+\delta)$-approximation for \maxcut, where $\delta = \delta(b)$ is some positive constant.

One of the components of our algorithm is a weak local
graph partitioning procedure that may be of independent interest. Given a starting vertex $i$ and a conductance parameter $\phi$, unless a random walk of length $\ell = O(\log n)$ starting from $i$ mixes rapidly (in terms of $\phi$ and $\ell$), we can find a cut of conductance at most $\phi$ close to the vertex. The work done per vertex found in the cut is sublinear in $n$.
\end{abstract}
\end{titlepage}

\section{Introduction}

The problem of finding the maximum cut of a graph is a classical combinatorial optimization problem. Given a graph $G = (V, E)$, with weights $w_{ij}$ on edges $\{i, j\}$, the problem is to partition the vertex set $V$ into two sets $L$ and $R$ to maximize the weight of \emph{cut edges} (these have one endpoint in $L$ and the other in $R$). The value of a cut is the total weight of cut edges divided by the total weight. The largest possible value of this is $\maxcut(G)$.
%
The problem of computing $\maxcut(G)$ was one of Karp's original NP-complete problems~\cite{Kar75}.

Therefore, polynomial-time approximation algorithms for \maxcut were sought out, that would provide a cut with value at least $\alpha \maxcut(G)$,
for some fixed constant $\alpha > 0$. It is easy to show that a random cut
gives a $0.5$-approximation for the \maxcut. This was the best known
for decades, until the seminal paper on semi-definite programming (SDP) by Goemans and Williamson~\cite{GW95}. They gave a $0.878\ldots$-approximation algorithm, which is optimal for polynomial time algorithms under the Unique Games Conjecture~\cite{Kho02,KKMO04}. Arora and Kale~\cite{AK07} gave an efficient near-linear-time implementation of the SDP algorithm for \maxcut\footnote{This was initially only proved for graphs in which the ratio of maximum to average degree was bounded by a polylogarithmic factor, but a linear-time reduction due to Trevisan~\cite{Tre09} converts any arbitrary graph to this case.}.

In spite of the fact that efficient, possibly optimal, approximation algorithms are known, there is a lot of interest in understanding what techniques are required to improve the $0.5$-approximation factor. By ``improve", we mean a ratio of the form $0.5+\delta$, for some constant $\delta > 0$. The powerful technique of Linear Programming (LP) relaxations fails to improve the $0.5$ factor. Even the use of strong LP-hierarchies to tighten relaxations does not help~\cite{dlVKM07,STT07}. Recently, Trevisan~\cite{Tre09} showed for the first time that a technique weaker than SDP relaxations can beat the $0.5$-factor. He showed that the eigenvector corresponding to the smallest eigenvalue of the adjacency matrix can be used to approximate the \maxcut to factor of $0.531$.
Soto~\cite{Sot09} gave an improved analysis of the same algorithm that provides a better approximation factor of $0.6142$. The running time\footnote{In this paper, we use the $\otilde$ notation to suppress dependence on polylogarithmic factors.} of this algorithm is $\otilde(n^2)$.

All the previous algorithms that obtain an approximation factor better than $0.5$ are not ``combinatorial", in the sense that they all involve numerical matrix computations such as eigenvector computations and matrix exponentiations. It was not known whether combinatorial algorithms can beat the $0.5$ factor, and indeed, this has been explicitly posed as an open problem
by Trevisan~\cite{Tre09}. Combinatorial algorithms are appealing because they exploit deeper insight into the combinatorial structure of the problem, and because they can usually be implemented easily and efficiently, typically without numerical round-off issues.

\subsection{Our contributions}

\begin{figure}[htbp] \centering
    \includegraphics[width=7in]{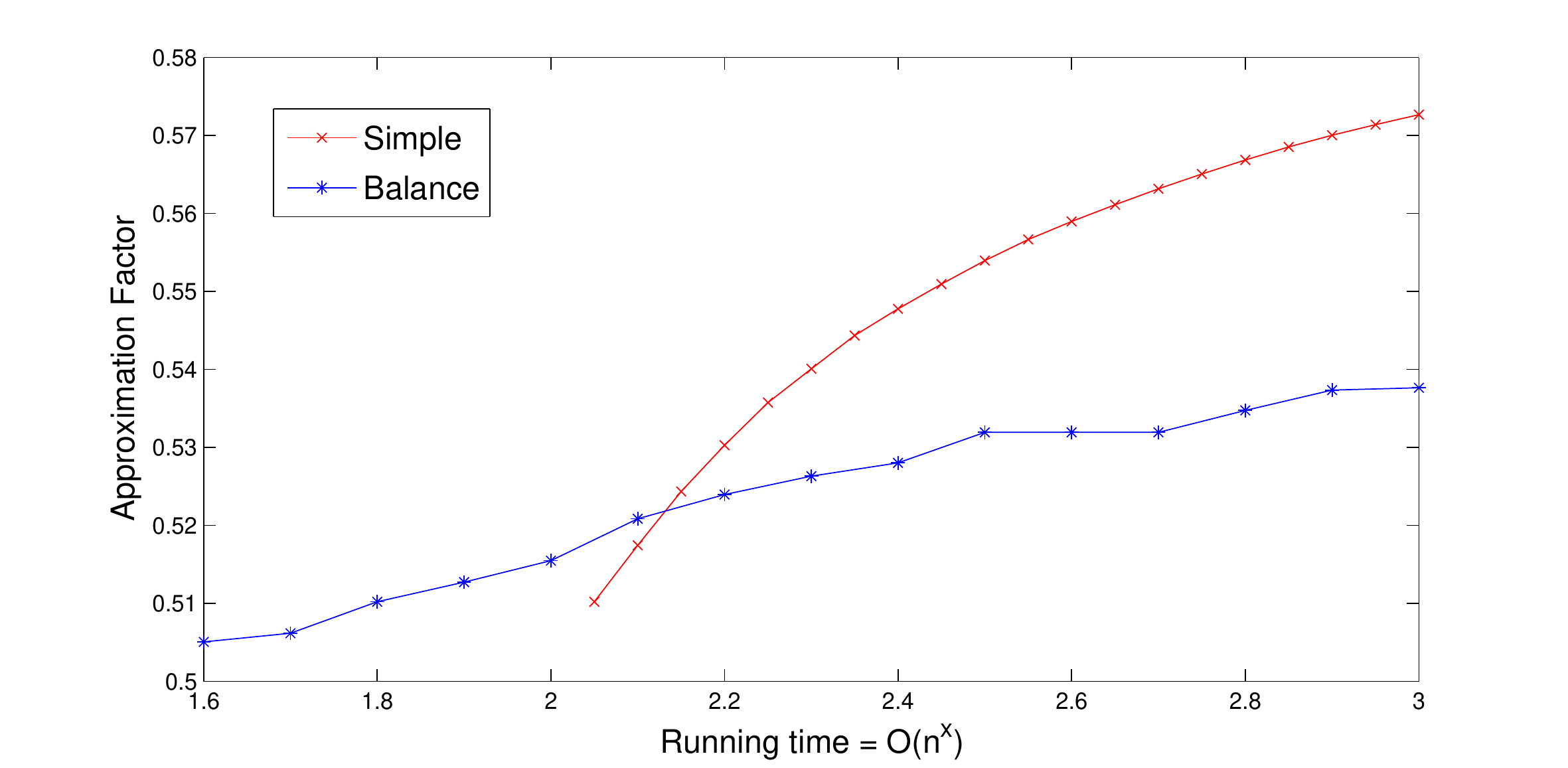} \caption{Running Time/Approximation Ratio Tradeoff Curve for \Simple and \Balance. \Simple
    needs running time $\otilde(n^{2+\mu})$ and \Balance needs running time $\otilde(n^{1.5+\mu)}$, for any constant $\mu > 0$. The approximation ratio for \Simple is from Lemma~\ref{lem:simple}, and that for \Balance is from Claim~\ref{clm:lp}.}
    \label{fig:simple-balance}
\end{figure}

1. In this paper, we achieve this goal of a combinatorial approximation
algorithm for \maxcut. We analyze a very natural, simple, and combinatorial
heuristic for finding the \maxcut of a graph, and show that it actually manages
to find a cut with an approximation factor strictly greater than $0.5$. In fact, we really
have a suite of algorithms:
\begin{theorem} \label{thm:main}
For any constant $b > 1.5$, there is a combinatorial algorithm
that runs in $\otilde(n^b)$ time and provides
an approximation factor that is a constant greater than $0.5$.
\end{theorem}
%
The running time/approximation factor tradeoff curve is
shown in Figure~\ref{fig:simple-balance}. A few representative numbers: in
$\otilde(n^{1.6})$, $\otilde(n^2)$, and $\otilde(n^3)$ times, we can get
approximation factors of $0.5051$, $0.5155$, and $0.5727$ respectively. As $b$
becomes large, this converges to the ratio of Trevisan's algorithm.


2. Even though the core of our algorithm is completely combinatorial, relying only on simple random walks and integer operations, the analysis of the algorithm is based on spectral methods. We obtain a combinatorial version of Trevisan's algorithm by showing two key facts: (a) the ``flipping signs'' random walks we use corresponds to running the power method on the graph Laplacian, and (b) a random starting vertex yields a good starting vector for the power method with constant probability. These two facts replace numerical matrix computations with the combinatorial problem of estimating certain probabilities, which can be done effectively by sampling and concentration bounds. This also allows improved running times since we can selectively find portions of the graph and classify them.

3. A direct application of the partitioning procedure yields an algorithm whose running time is $\otilde(n^{2+\mu})$. To design the sub-quadratic time algorithm, we have to ensure that the random walks in the algorithm mix rapidly. To do this, we design a sort of a local graph partitioning algorithm of independent interest based on simple random walks of logarithmic length. Given a starting vertex $i$,
either it finds a low conductance cut or certifies that the random walk from $i$ has somewhat mixed, in the sense that the ratio of the probability of hitting any vertex $j$ to its probability in the stationary distribution is bounded. The work done per vertex output in the cut is sublinear in $n$. The precise statement is given in Theorem~\ref{thm:max}.
Previous local partitioning algorithms~\cite{ST04, ACL06, AL08} are more efficient than our procedure, but
can only output a low conductance cut, if the actual conductance of some set containing $i$ is $O(1/\log n)$.
In this paper, we need to be able to find low conductance cuts in more general settings, even if there is no cut of conductance of $O(1/\log n)$, and hence the previous algorithms are unsuitable for our purposes.



\subsection{Related work}

Trevisan~\cite{Tre05} also uses random walks to give approximation algorithms for \maxcut (as a special case of unique games), although the algorithm only deals with the case when \maxcut is $1-O(1/\poly(\log n))$.
The property tester for bipartiteness in sparse graphs
by Goldreich and Ron~\cite{GR99} is a sublinear time procedure that uses
random walks to distinguish
graphs where $\maxcut =1$ from $\maxcut \leq 1 - \epsilon$.
The algorithm, however, does not actually give an approximation to \maxcut.
There is a similarity in flavor to Dinur's proof of the PCP theorem~\cite{Din05},
which uses
random walks and majority votes for gap amplification of CSPs.
Our algorithm might be seen as some kind of belief
propagation, where messages about labels are passed around.

For the special case of cubic and maximum
degree $3$ graphs, there has been a study of combinatorial
algorithms for \maxcut~\cite{BL86,ELZ04,BT08}. These
are based on graph theoretic properites and very different
from our algorithms. Combinatorial algorithms for CSP
(constraint satisfaction problems) based on LP
relaxations have been studied in~\cite{DDFGMP03}.

\section{Algorithm Overview and Intuition}

Let us revisit the greedy algorithm. We currently have a partial
cut, where some subset $S$ of the vertices have been classified (placed
in either side of the cut). We take a new vertex $i \notin S$ and look at the
edges of $i$ incident to $S$. In some sense, each such edge provides
a ``vote" telling $i$ where to go.
Suppose there is such an edge $(i,j)$, such that $j \in R$.
Since we want to cut edges, this edge tells $i$ to be placed in $L$.
We place $i$ accordingly to a majority vote, and hence the $0.5$ factor.

Can we take that idea further, and improve on the $0.5$ factor?
Suppose we fix a source vertex $i$ and
try to classify vertices with respect to the source.
Instead of just looking at edges (or paths of length $1$),
let us look at longer paths. Suppose we choose a length
$\ell$ from some nice distribution (say, a binomial
distribution with a small expectation)
and consider paths of length $\ell$ from $i$.
If there are many more even length paths to $j$ than odd length
paths, we put $j$ in $L$, otherwise in $R$.
This gives a partition of vertices that we can reach,
and suggests an algorithm based on random walks.
We hope to estimate the odd versus even length probabilities
through random walks from $i$. This is a very natural
idea and elegantly extends the greedy approach. Rather
surprisingly, we show that this can be used to beat
the $0.5$ factor by a constant.

One of the main challenges is to show that we do not
need too many walks to distinguish these various
probabilities. We also need to choose our length
carefully. If it is too long, then the odd and even path
probabilities may become too close to each other.
If it is too short, then it may not be enough to
get sufficient information to beat the greedy
approach.

Suppose the algorithm detects that the probability of going from vertices $i$ to $j$ by an odd length path is significantly higher than an even length path. That suggests that we can be fairly confident that $i$ and $j$ should be on different sides of the cut.
This constitutes the core of our algorithm, \Threshold. This algorithm classifies some vertices as lying on ``odd'' or ``even'' sides of the cut based on which probability (odd or even length paths) is significantly higher than the other.  Significance is decided by a threshold that is a parameter to the algorithm. We show a connection between this algorithm and Trevisan's, and then we adapt his (and Soto's) analysis to show that one can choose the threshold carefully so that amount of work done per classified vertex is bounded, and the number of uncut edges is small. The search for the right threshold is done by the \Find algorithm.

Now, this procedure leaves some vertices unclassified, because no probability is significantly larger than the other. We can simply recurse on the unclassified vertices, as long as the the cut we obtain is better than the trivial $0.5$ approximate cut. This constitutes the \Simple algorithm. The analysis of this algorithm shows that we can bound the work done per vertex is at most $\otilde(n^{1+\mu})$ for any constant $\mu > 0$, and thus the overall running time becomes $\otilde(n^{2+\mu})$. This almost matches the running time of Trevisan's algorithm, which runs in $\otilde(n^2)$ time.


To obtain a sub-quadratic running time, we need to do a more careful analysis of the random walks involved. If the random walks do not mix rapidly, or, in other words, tend to remain within a small portion of the graph, then we end up classifying only a small number of vertices, even if we run a large number of these random walks. This is why we get the $\otilde(n^{1+\mu})$ work per vertex ratio.

But in this case, we can exploit the connection between fast mixing and high conductance~\cite{sinclair, M89, LS90} to conclude that there must be a low conductance cut which accounts for the slow mixing rate. To make this algorithmic, we design a local graph partitioning algorithm based on the same random walks as earlier. This algorithm, \CutOrBound, finds a cut of (low) constant conductance if the walks do not mix, and takes only around $\otilde(n^{0.5 + \mu})$ time, for any constant $\mu > 0$, per vertex found in the cut. Now, we can remove this low conductance set, and run \Simple on
the induced subgraph. In the remaining piece, we recurse. Finally, we combine the cuts found randomly.
This may leave up to half of the edges in the low conductance cut uncut, but that is only a small constant fraction of the total number of edges overall. This constitutes the \Balance algorithm. We show that we spend only $\otilde(n^{0.5 + \mu})$ time for every classified vertex, which leads to a $\otilde(n^{1.5 + \mu})$ overall running time.

All of these algorithms are combinatorial: they only need random selection of outgoing edges, simple arithmetic operations, and comparisons. Although the analysis is technically involved, the algorithms themselves are simple and easily implementable.

\section{The Threshold Cut} \label{sec:cut}

We now describe our core random walk based procedure to partition vertices.
Some notation first.
The graph $G$ will have $n$ vertices. All our algorithms will be based on lazy
random walks on $G$ with self-loop probability $1/2$. We define these walks now. Fix a
length $\ell = O(\log n)$. At each step in the random walk, if we are currently
at vertex $j$, then in the next step we stay at $j$ with probability $1/2$.
With the remaining probability ($1/2$), we choose a random incident edge $\{j,
k\}$ with probability proportional to $w_{jk}$ and move to $k$. Thus the edge
$\{j, k\}$ is chosen with overall probability $w_{jk}/2d_j$, where $d_j =
\sum_{\{j, k\} \in E} w_{jk}$ is the (weighted) degree of vertex $j$. Let
$\Delta$ be an upper bound on the maximum degree. By a linear time reduction of
Trevisan~\cite{Tre01,Tre09}, it suffices to solve \maxcut on graphs\footnote{We
can think of these as unweighted multigraphs.} where $\Delta = \poly(\log n)$.
We set $m$ to be sum of weighted degrees, so $m := \sum_j d_j$. We note that by
Trevisan's reduction, $m = \otilde(n)$, and thus running times stated in terms
of $m$ translate directly to the same polynomial in $n$.

The random walk described above is equivalent to flipping an unbiased coin $\ell$ times, and running a simple (non-lazy) random walk for $h$ steps, where $h$ is the number of heads seen. At each step of this simple random walk, an outgoing edge is chosen with probability proportional to its weight. We call $h$ the {\em hop-length} of the random walk, and we call a walk odd or even based on the parity of $h$.

We will denote the two sides of the cut by $L$ and $R$. The parameters $\epsilon$ and $\mu$ are fixed throughout this section, and should
be considered as constants. We will choose the length $\ell$ of the
walk to be $\mu(\ln (4m/\delta^2))/[2(\delta + \epsilon)]$
(the reason for this choice will be explained later). We will assume that $\gamma$ and $\delta$ are arbitrarily small constants. The procedure {\Threshold} takes as input a \emph{threshold} $t$, and puts \emph{some} vertices in one of two sets, $Even$ and $Odd$, that are assumed to be global variables (i.e. different calls to \Threshold update the same sets). We call vertices $j \in Even \cup Odd$ {\em classified}. Once classified, a vertex is never re-classified. We perform a series of random walks to decide this. The number of walks will be a function of this threshold $w(t)$.
We will specify this function later.
\begin{center}
\fbox{\begin{minipage}{\columnwidth} \Threshold \ \ \ \ \ {\bf Input:} Graph $G = (V, E)$. \ \ \ \ {\bf Parameters:} Starting vertex $i$, threshold $t$.
\begin{enumerate*}
    \item Perform $w(t)$ walks of length $\ell$ from $i$.
    \item For every vertex $j$ that is not classified:
    \begin{enumerate*} 
	    \item Let $\qbi(j) := \frac{1}{d_jw(t)}(\#\textrm{\{even walks ending at $j$\}} - \#\textrm{\{odd walks ending at $j$\}})$.
	    \item If $\qbi(j) > t$, put $j$ in set $Even$. If $\qbi(j) < -t$, put it in set $Odd$.
		\end{enumerate*}
\end{enumerate*}
\end{minipage}}
\end{center}
%

We normalize the difference of the number of even and odd walks by $d_j$ to
account for differences in degrees. This accounts for the fact that the
stationary probability of the random walk at $j$ is proportional to $d_j$. For
the same reason, when we say ``vertex chosen at random" we will mean choosing a
vertex $i$ with probability proportional to $d_i$. We now need some definitions.
\begin{Def}[Work-to-output ratio.]
Let $\mathcal{A}$ be an algorithm that, in time $T$, classifies $k$ vertices (into the sets $Even$ or $Odd$). Then the work-to-output ratio of $\mathcal{A}$ is defined to be $\frac{T}{k}$.
\end{Def}
\begin{Def}[\good, \cross, \inc, \cut.] \label{def:cut}
Given two sets of vertices $A$ and $B$, let $\good(A,B)$ be the total weight of edges that have one endpoint in $A$ and the other in $B$. Let $\cross(A,B)$ be the total weight of edges with only one endpoint in $A \cup B$. Let $\inc(A,B)$ be the total weight of edges incident on $A \cup B$. We set $\cut(A,B) := \good(A,B) + \cross(A,B)/2$.
\end{Def}

Suppose we either put all the vertices in $Even$ in $L$ or $R$, and the vertices in
$Odd$ in $R$ or $L$ respectively, retaining whichever assignment cuts more
edges. Then the number of edges cut is at least $\cut(Even, Odd)$.


\begin{Def}[$\alpha$, $w(t)$, $\sigma$, $\soto(\sigma)$.] \label{def:alpha-sigma-etc}
\begin{enumerate}
    \item For every vertex $j$, let $p^\ell_j$ be the probability of reaching $j$
    starting from $i$ with an $\ell$-length lazy random walk.
    Let $\alpha$ be an upper bound on $\max_j \frac{p^\ell_j}{d_j}$.

    \item Define $w(t) := \frac{\kappa \ln(n)\max\{\alpha, t\}}{t^2}$, for a large enough constant $\kappa$.

    \item Define $\sigma := 1 - (1 - \epsilon)^{1 + \frac{1}{\mu}} - o(1)$, where the $o(1)$ term can be made as small as we please by setting $\delta, \gamma$ to be sufficiently small constants.

    \item Define the function $\soto(\sigma)$ (c.f. \cite{Sot09})
    as follows: here $\sigma_0 = 0.22815\ldots$ is a fixed constant. If $\sigma
    > 1/3$, then  $\soto(\sigma) = 0.5$. If $\sigma_0 < \sigma \leq 1/3$, then
    $\soto(\sigma) =  \frac{-1+\sqrt{4\sigma^2 - 8\sigma + 5}}{2(1-\sigma)}$.
    Otherwise, $\soto(\sigma) = \frac{1}{1+2\sqrt{\sigma(1-\sigma)}}$.
\end{enumerate}
\end{Def}

The parameter $\alpha$ measures how far the walk is from mixing, because the stationary probability of $j$ is proportional to $d_j$. The function $\soto(\sigma) > 0.5$ when $\sigma < 1/3$, and this leads to an approximation factor greater than $0.5$. Now we state our main performance bound for \Threshold.
\begin{Lem} \label{lem:threshold-perf}
Suppose $\maxcut \geq 1 - \epsilon$. Then, there is a threshold $t$ such that with constant probability over the choice of a starting vertex $i$ chosen at random, the following holds. The procedure $\Threshold(i, t)$ outputs sets $Even$ and $Odd$ such that
$\cut(Even,Odd)\ \geq\ \soto(\sigma)\inc(Even,Odd)$. Furthermore, the work-to-output ratio is bounded by $\otilde(\alpha \Delta m^{1+\mu} + 1/\alpha)$.
\end{Lem}

The main procedure of this section, {\Find}, is just an algorithmic version of
the existential result of Lemma~\ref{lem:threshold-perf}.


%
\begin{center}
\fbox{\begin{minipage}{\columnwidth} {\Find} \ \ \ \ \ {\bf Input:} Graph $G = (V, E)$. \ \ \ \ {\bf Parameters:} Starting vertex $i$
\begin{enumerate*}
    \item Initialize sets $Even$ and $Odd$ to empty sets.
    \item For $t_r = (1-\gamma)^r$, for $r = 0,1,2,\ldots$, as long as $t_{r} \geq \gamma/ m^{1+\mu/2}$. 
    \begin{enumerate*}
			\item Run \Threshold$(i,t_r)$.
	    \item If $\cut(Even,Odd) \geq \soto(\sigma)\inc(Even,Odd)$
	    and $|Even \cup Odd| \geq (\Delta t_r^2n^{1+\mu}\log n)^{-1}$, output $Even$ and $Odd$. Otherwise go to the next threshold.
	  \end{enumerate*}
	  \item Output FAIL.
\end{enumerate*}
\end{minipage}}
\end{center}
%
%
We are now ready to state the performance bounds for \Find.

\begin{Lem} \label{lem:find} Suppose $\maxcut \geq 1 -\epsilon$. Let $i$ be chosen at random.
With constant probability over the choice of $i$ and the randomness of
$\Find(i)$, the procedure $\Find(i)$ succeeds and has a work to output ratio of
$\otilde(\alpha \Delta m^{1+\mu} + 1/\alpha)$. Furthermore, regardless of the
value of \maxcut or the choice of $i$, the worst-case running time of
$\Find(i)$ is $\otilde(\alpha \Delta m^{2+\mu})$.
\end{Lem}

The proofs of Lemmas~\ref{lem:threshold-perf} and~\ref{lem:find} use
results from Trevisan's and Soto's analyses~\cite{Tre09,Sot09}.
The vectors we consider will always be $n$-dimensional,
and should be thought of as an assignment of values to each of the $n$ vertices
in $G$.
Previous analyses rest on the fact that a vector
that has a large Rayleigh quotient
(with respect to the graph Laplacian\footnote{For a vector $x$
and matrix $M$, the Rayleigh quotient is $\frac{x^\top M x}{x^\top x}$.}) can be
used to find good cuts. Call such a vector ``good".
These analyses show that partitioning vertices by \emph{thresholding} over
a good vector $x$ yields a good cut. This means that for some threshold $t$,
vertices $j$ with $x(j) > t$ are placed in $L$ and those with $x(j) < -t$
are placed in $R$.
We would like to show that {\Threshold} is
essentially performing such a thresholding on some good vector. We will construct a vector,
somewhat like a distribution, related to {\Threshold}, and show that it is
good. This requires an involved spectral
analysis. This is formalized in Lemma~\ref{lem:good-q}. With this
in place, we use concentration inequalities and an adaptation of the techniques
in~\cite{Sot09} to connect thresholding to the cuts looked at by {\Find}.
We first state Lemma~\ref{lem:good-q}. Then we will show how
to prove Lemmas~\ref{lem:threshold-perf} and~\ref{lem:find} using Lemma~\ref{lem:good-q}.
This is rather involved, but intuitively should be fairly clear.
It mainly requires understanding of
the random process that {\Threshold} uses to classify vertices.

We need some
definitions. Let $A$ be the (weighted) adjacency matrix of $G$ and $d_i$
be the degree of vertex $i$.  The (normalized) Laplacian of the graph is $\lap
= I - \Dhalfinv A\Dhalfinv$. Here $D$ is the matrix where $D_{ii} = d_i$ and
$D_{ij} = 0$ (for $i \neq j$). For a vector $x$ and coordinate/vertex $j$, we use $x(j)$
to denote the $j$th coordinate of $x$ (we do \emph{not} use subscripts
for coordinates of vectors).
In~\cite{Tre09} and~\cite{Sot09}, it was shown
that vectors that have high Rayleigh quotients with $\lap$
can be used to get a
partition that cuts significant number of edges. Given a vector $y$, let us do
a simple rounding to get partition vertices. We define the sets $ P(y,t) = \{
j\ |\ y(j) \geq t\}$ and $N(y, t) = \{j\ |\ y(j) \leq -t\} $. We refer to
rounding of this form as \emph{tripartitions}, since we divide the vertices
into three sets. The following lemma, which is Lemma 4.2 from~\cite{Sot09}, an
improvement of the analysis in~\cite{Tre09}, shows that this tripartition cuts
many edges for some threshold:

\begin{Lem}\label{lem:soto} (\cite{Sot09}) Suppose $x^\top \lap x \geq 2(1-\sigma) \|x\|^2$.
Let $y = \Dhalfinv x$.
Then, for some $t$ (called \emph{good}),
$\cut(P(y,t),N(y,t)) \geq \soto(\sigma) \inc(P(y,t),N(y,t))$.
\end{Lem}

The algorithm of Trevisan is the following: compute the top eigenvector $x$ of $\lap$ (approximately), compute $y = \Dhalfinv x$, and find a good threshold $t$ and the corresponding sets $P(y, t), N(y, t)$. Assign $P(y, t)$ or $N(y, t)$ to $L$ and $R$ (or vice-versa, depending on which assignment cuts more edges), and recurse on the remaining unclassified vertices.

The algorithms of this paper essentially mimic this process, except that
instead of computing the top eigenvector, we use random walks. We establish a
connection between random walks and the power method to compute the top
eigenvector. Let $p_{i,j}^h$ be the probability that a length $\ell$ (remember that this
is fixed) lazy
random walk from $i$ reaches $j$ with hop-length $h$. Then define the vector
$\qi$ as follows: the $j$th coordinate of $\qi$ is $ \qi(j) :=
\frac{1}{\sqrt{d_j}}\left( \sum_{h \ \textrm{even}} p_{i,j}^h - \sum_{h \
\textrm{odd}} p_{i,j}^h \right) = \frac{1}{\sqrt{d_j}}\sum_{h = 0}^\ell (-1)^h
p_{i,j}^h.$

Note that {\Threshold} is essentially computing an estimate $\qbi(j)$
of $\qi(j)/\sqrt{d_j}$. For convenience, we will denote $\Dhalfinv \qi$ by $\halfqi$.
This is the main lemma of this section.
%
\begin{Lem}\label{lem:good-q}
Let $\delta > 0$ be a sufficiently small constant, and $\mu > 0$ be
a (constant) parameter.
If $\ell = \mu(\ln (4m/\delta^2))/[2(\delta + \epsilon')]$, where $\epsilon' = -\ln(1-\epsilon)$, then
with constant probability over the choice of $i$,
$\| \qi \|^2 = \Omega(1/m^{1+\mu})$, and
\begin{equation} \label{eq:lap-q}
{\qi}^\top \lap {\qi}\ \geq\ 2e^{-(2 + \frac{1}{\mu})\delta}(1 - \epsilon)^{1 + \frac{1}{\mu}}\|{\qi}\|^2,
\end{equation}
\end{Lem}

Although this not at all straightforward, it appears that Lemma~\ref{lem:good-q}
with Lemma~\ref{lem:soto} essentially proves Lemma~\ref{lem:threshold-perf}.
To ease the flow of the paper, we defer these arguments to Section~\ref{sec:thresh}.

Lemma~\ref{lem:good-q} is proved in two parts. In the first, we establish a connection
between the random walks we perform and running the power method on the
Laplacian:
\begin{claim} \label{lem:lap-rw} Let $e_i$ be $i^\text{th}$ standard basis vector. Then, we
have
$\qi =\ \frac{1}{2^\ell} \lap^\ell \left(\frac{1}{\sqrt{d_i}}e_i\right)$.
\end{claim}
\begin{myproof} Note that $\lap^\ell = (I - \Dhalfinv A \Dhalfinv)^\ell = (\Dhalfinv(I -
AD^{-1})\Dhalf)^\ell = \Dhalfinv(I - AD^{-1})^\ell\Dhalf$. Hence,
\begin{eqnarray*}
\lap^\ell\left(\frac{1}{\sqrt{d_i}}e_i\right) & = & \Dhalfinv(I - AD^{-1})^\ell e_i \\
& = & 2^\ell \Dhalfinv\sum_{h=0}^\ell (-1)^h{\ell \choose h} \left(\frac{1}{2}\right)^{\ell - h} \left(\frac{1}{2}AD^{-1}\right)^h e_i  = 2^\ell \qi
\end{eqnarray*}
%
The last equality follows because the vector ${\ell \choose h}
\left(\frac{1}{2}\right)^{\ell - h} \left(\frac{1}{2}AD^{-1}\right)^h e_i$ is
the vector of probabilities of reaching different vertices starting from $i$ in a walk of length $\ell$ with hop-length exactly $h$. We also used the
facts that $\Dhalf \frac{1}{\sqrt{d_i}} e_i = e_i$ and $\Dhalfinv e_j =
\frac{1}{\sqrt{d_j}} e_j$.
\end{myproof}\\

In the second part, we show that with constant probability, a randomly chosen starting vertex yields a good starting vector for the power method, i.e., the vector $\qi$ satisfies (\ref{eq:lap-q}). This will require a spectral analysis.  We need some notation first.
Let the eigenvalues of $\lap$ be $2 \geq \lambda_1 \geq \lambda_2 \geq \cdots
\lambda_n = 0$, and let the corresponding (unit) eigenvectors be $v_1, v_2,
\ldots v_n = \Dhalf \vec{1}_V$.
For a subset $S$ of vertices, define $\vol(S) = \sum_{i \in S} d_i$. Let $H =
\{k:\ \lambda_k \geq 2e^{-\delta}(1-\epsilon)\}$. Any vector $x$ can be
expressed in terms of the eigenvectors of $\lap$ as $x = \sum_k \alpha_k v_k$.
Define the norm $\|x\|_H\ =\ \sqrt{\sum_{k \in H} \alpha_k^2}$.

Let $(S, \bar{S})$ be the max-cut, where we use the convention $\bar{S} = V \setminus S$. Let $\vol(S) \leq \vol(V)/2$ and define
$s := \vol(S)$. Note that $m = \vol(V)$. Since the max-cut has size at least $(1-\epsilon)m/2$, we must have $s \geq (1-\epsilon)m/2$.
We set the vector $x = \Dhalf y$ where $y = \frac{1}{s}\vec{1}_S - \frac{1}{m}\vec{1}_V$
where $\vec{1}_S$ is the indicator vector for $S$. We will need some
preliminary claims before we can show (\ref{eq:lap-q}).

\begin{claim} \label{clm:hnorm-x} $\|x\|_H^2 \geq \delta/m$.
\end{claim}

\begin{myproof} We have
$$x^\top \lap x\ =\ \sum_{i,j \in E} (y(i) - y(j))^2\ =\ E(S, \bar{S}) \cdot \frac{1}{s^2}\ \geq\ \frac{(1-\epsilon)\cdot (m/2)}{s^2}\ \geq\
\frac{(1-\epsilon)m}{2s^2}$$ 
Now\footnote{This is easily seen using Pythagoras: since $\Dhalf
(\frac{1}{s}\vec{1}_S - \frac{1}{m}\vec{1}_V) \cdot \Dhalf \vec{1}_V = 0$. This
only uses the fact that $S \subseteq V$.},
$\|x\|^2 = \frac{1}{s} - \frac{1}{m}$.
Let $x = \sum_k \alpha_k v_k$ be the
representation of $x$ in the basis given by the $v_k$'s, and let $a :=
\|x\|_H^2$. Then we have $\|x\|^2 = \sum_k \alpha_k^2$, and
$$x^\top \lap x\ =\ \sum_k \lambda_k \alpha_k^2 \leq 2\sum_{k \in H} \alpha_k^2 +
2e^{-\delta}(1-\epsilon)\sum_{k \notin H} \alpha_k^2\ =\ 2a +
2e^{-\delta}(1-\epsilon)\left(\frac{1}{s} - \frac{1}{m} - a\right).$$
%

Combining the two bounds, and solving for $a$, we get the required bound for small enough $\delta$.
\end{myproof}


\begin{claim} \label{clm:hnorm-e}
With constant probability over the choice of $i$, $\| e_i \|_H^2 > \delta d_i/4m$.
\end{claim}
\begin{myproof} Let $T := \{i \in S:\ \|\frac{1}{\sqrt{d_i}}e_i\|^2_H <
\frac{\delta}{4m}\}$, and let $t = \vol(T)$. Our aim is to show that
$t$ is at most a constant fraction of $s$. For the sake of
contradiction, assume $t \geq (1-\theta)s$,
where $\theta = \delta(1 - \epsilon)/16$. Let $z = \Dhalf(\frac{1}{t}\vec{1}_T - \frac{1}{m}\vec{1})$.
We have
$$\|x - z\|_H^2\ \leq\ \|x - z\|^2\ =\ \frac{1}{t} - \frac{1}{s}\ \leq\
\frac{2\theta}{s}\ \leq\ \frac{4\theta}{(1-\epsilon)m}\ =\
\frac{\delta}{4m}$$
The second equality above uses the fact that $t \geq (1-\theta)s$ and $\theta < 1/2$.
The third inequality follows from $s \geq (1-\epsilon)m/2$.
By the triangle inequality and Claim~\ref{clm:hnorm-x}, we have
$$\|z\|_H\ \geq\ \|x\|_H - \|x - z\|_H\ \geq\ \sqrt{\frac{\delta}{m}} -
\sqrt{\frac{\delta}{4m}}\ =\ \sqrt{\frac{\delta}{4m}}$$
Now, we have $z = \sum_{i \in T} (\frac{d_i}{t})\Dhalf(\frac{1}{d_i}e_i - \frac{1}{m}\vec{1}_V)$,
so by Jensen's inequality, we get
$$\frac{\delta}{4m}\ \leq\ \|z\|_H^2\ \leq\ \sum_{i \in T}\frac{d_i}{t} \cdot
\left\|\Dhalf\left(\frac{1}{d_i}e_i - \frac{1}{m}\vec{1}_V\right)\right\|_H^2\ =\ \sum_{i \in T}\frac{d_i}{t} \cdot
\left\|\Dhalf\left(\frac{1}{d_i}e_i\right)\right\|_H^2\ <\ \frac{\delta}{4m},$$ a contradiction.
The equality in the chain above holds because $\Dhalf \vec{1}_V$ has no component along
the eigenvectors corresponding to $H$ (this is an eigenvector itself,
with eigenvalue $0$).

Thus the set $S \setminus T$ has volume at least $\theta s \geq
\delta(1-\epsilon)^2m/32$. Note that the sampling process, which chooses
the initial vertex of the random walk by choosing a random edge and choosing a
random end-point $i$ of it, hits some vertex in $S \setminus T$ with probability
at least $\frac{\vol(S \setminus T)}{\vol(V)}\ \geq\
\delta(1-\epsilon)^2/32$, i.e. constant probability.
\end{myproof}
\medskip

At this point, standard calculations for the power method imply Lemma~\ref{lem:good-q}. \\

\begin{myproof}(Of Lemma~\ref{lem:good-q})
From Claim~\ref{clm:hnorm-e}, with constant probability $\|e_i\|_H^2 \geq \delta d_i/4m$. Let us assume this is case.

For convenience, define $\beta = \frac{(\delta + \epsilon')}{\mu}$, so that the number of walks is $\ell = \frac{\ln(4m/\delta^2)}{2\beta}$.  Now let $H' = \{i: \lambda_i \geq 2e^{-(\delta + \beta)}(1-\epsilon)\}$.
Write $\frac{1}{\sqrt{d_i}}e_i$ in terms of the $v_k$'s as
$\frac{1}{\sqrt{d_i}}e_i = \sum_k \alpha_k v_k$. Let $\qhi = \lap^\ell
\frac{1}{\sqrt{d_i}}e_i = \sum_k \alpha_k \lambda_k^\ell v_k$. Note that $\qi = \frac{1}{2^\ell}\qhi$. Then we have
$$\qhi^\top \lap \qhi\ =\ \sum_k \alpha_k^2 \lambda_k^{2\ell+1}\ \geq\ \sum_{k
\in H'} \alpha_k^2 \lambda_k^{2\ell} \cdot
2e^{-(\delta + \beta)}(1-\epsilon)$$
and
$$\|\qhi\|^2 = \sum_k \alpha_k^2
\lambda_k^{2\ell} = \sum_{k \in H'} \alpha_k^2 \lambda_k^{2\ell}\left[1 +
\frac{\sum_{k \notin H'} \alpha_k^2 \lambda_k^{2\ell}}{\sum_{k \in H'}
\alpha_k^2 \lambda_k^{2\ell}}\right].$$
We have
$$\frac{\sum_{k \notin H'} \alpha_k^2 \lambda_k^{2\ell}}{\sum_{k \in H'} \alpha_k^2
\lambda_k^{2\ell}}\ \leq\ \frac{\sum_{k \notin H'} \alpha_k^2 \lambda_k^{2\ell}}{\sum_{k \in H} \alpha_k^2
\lambda_k^{2\ell}}\ \leq\ \frac{(2e^{-(\delta + \beta)}(1-\epsilon))^{2\ell}}{\frac{\delta}{4m}(2e^{-\delta}(1-\epsilon))^{2\ell}}\ =\ \frac{4m e^{-2\beta\ell}}{\delta}.$$
Thus,
$$\frac{\qhi^\top \lap \qhi}{\|\qhi\|^2}\ \geq\ \frac{2e^{-(\delta + \beta)}(1-\epsilon)}{1 +
\frac{4m e^{-2\beta\ell}}{\delta}}$$
Observe that $\qhi$ is just a scaled version of $\qi$, so we
can replace $\qhi$ by $\qi$ above. For the denominator in the right,
we would like to set that to be $e^\delta$. Choosing
$\ell = \frac{\ln(4m/\delta^2)}{2\beta}$, we get
$$ {\qi}^\top \lap {\qi}\ \geq\ 2e^{-(2\delta + \beta)}(1 - \epsilon)\|{\qi}\|^2\ =\ 2e^{-(2 + \frac{1}{\mu})\delta}(1-\epsilon)^{1 + \frac{1}{\mu}}\|\qi\|^2.$$

Since $\|e_i\|_H^2 \geq \delta d_i/4m$, we have
$$\sum_{k \in H} \alpha_k^2\ =\ \left\|\frac{1}{\sqrt{d_i}}e_i\right\|_H^2\ \geq\ \frac{\delta}{4m}.$$

This implies
$$\|q_i\|^2\ =\ \frac{1}{2^{2\ell}}\|\qhi\|^2\ =\ \frac{1}{2^{2\ell}}\sum_k \alpha_k^2 \lambda_k^{2\ell}\ \geq\ \frac{1}{2^{2\ell}}\sum_{k \in H} \alpha_k^2\lambda_k^{2\ell}.$$

By definition, for all $k \in H$, $\lambda_k \geq 2e^{-\delta}(1-\epsilon)$.
This gives a lower bound on the rate of decay of these coefficients,
as the walk progresses.
$$ \|q_i\|^2\ \geq\ \frac{1}{2^{2\ell}}\sum_{k \in H}\alpha_k^2 (2e^{-\delta}(1-\epsilon))^{2\ell}\ \geq\ \frac{\delta}{4m} e^{-2\delta \ell}(1-\epsilon)^{2\ell}\ =\ \Omega\left(\frac{1}{4m^{1+\mu}}\right),$$
by our choice of $\ell = \frac{\ln(4m/\delta^2)}{2\beta}$.
\end{myproof}

\subsection{Proofs of Lemmas~\ref{lem:threshold-perf} and~\ref{lem:find}} \label{sec:thresh}

Both Lemma~\ref{lem:find} and
Lemma~\ref{lem:threshold-perf} follow directly
from the following statement.

\begin{Lem} \label{lem:good-thresh} Let $w(t) = (c'\ln^2/\gamma^2)(\alpha/t^2)$, where
$c'$ is a sufficiently large constant. Let $C_r$ denote the set of vertices
classified by \Threshold$(i,t_r)$.
The following hold with constant probability
over the choice of $i$ and the randomness of {\Threshold}.
There exists a threshold $t_r = (1-\gamma)^r$ such
that $\sum_{j \in C_r} d_j = \Omega((t_r^2m^{1+\mu}\log n)^{-1})$.
Also, the tripartition generated satisfies Step 2(b) of {\Find}.
\end{Lem}

In this section, we will prove this lemma. But first,
we show how this implies Lemmas~\ref{lem:threshold-perf}
and~\ref{lem:find}.\\

\begin{myproof} (of Lemma~\ref{lem:threshold-perf}) We take
the threshold $t_r$ given by Lemma~\ref{lem:good-thresh}.
Since it satisfies Step 2(b) of {\Find},
$\cut(Even,Odd) \geq f(\sigma)\inc(Even,Odd)$. To see the work
to output ratio, observe that the work done is $\otilde(w(t_r))
= \otilde(\max(\alpha,t_r)/t_r^2)$. It is convenient
to write this as $\otilde(\alpha/t_r^2 + 1/\alpha)$.
The output is $C_r$. We have
$$ \Delta |C_r| \geq \sum_{j \in C_r} d_j = \Omega(\frac{1}{t_r^2m^{1+\mu}\log n}) $$
The output is at least $1$. Therefore, the work per
output is at most $\otilde(\alpha \Delta m^{1+\mu} + 1/\alpha)$.
\end{myproof}\\

\begin{myproof} (of Lemma~\ref{lem:find}) The running time when there
is failure is easy to see. The running time upto round $r$
is $\otilde(\sum_{j \leq r} \max(\alpha,t_j)/t_j^2) = \otilde(\alpha/t_r^2 + 1/\alpha)$.
Since $r^* = 1/n^{1+\mu/2}$ and $\alpha \leq 1/n^2$, we get the desired bound.
By Lemma~\ref{lem:good-thresh}, we know that {\Find} succeeds with high probability.
We have some round $r$ where {\Find} will
terminate (satisfying the conditions of Step 2(b)).
The work to output ratio analysis is the same as the previous
proof, and is at most $\otilde(\alpha \Delta m^{1+\mu} + 1/\alpha)$.
\end{myproof}\\
%

We will first need some auxilliary claims that
will help us prove Lemma~\ref{lem:good-thresh}.
The first step
is the use concentration inequalities to bound
the number of walks required to get coordinates of $\halfqi$.
As mentioned before, we designate the coordinates of $\qi$ by $\qi(j)$. The $p_i$
vector is the probability vector of the random walk (without charges)
for $\ell$ steps. In other words:
$$\halfqi(j) := (\pr[\text{Walk from $i$ reaches $j$ in even path}] - \pr[\text{Walk from $i$ reaches $j$ in odd path}])/d_j$$
and
$$p_i(j) := \pr[\text{Walk from $i$ reaches $j$ in even path}] + \pr[\text{Walk from $i$ reaches $j$ in odd path}]$$
This clearly shows that the random walks performed
by \Threshold{} are being used to estimate coordinates of $\qi$.
The following claim shows how many $w$ walks are required
to get good a approximation of coordinates $\qi$.

\begin{claim} \label{clm:est-qi} Suppose $w$ walks are performed. Let
$c$ be a sufficiently large constant and $1/\ln n < \gamma < 1$. The following
hold with probability at least $> 1 - n^{-4}$.
\begin{itemize}
	\item If $w \geq (c\ln n/\gamma^2) (\max(\alpha,t)/t^2)$, then we can get
	an estimate $\qbi(j)$ such that $\sqrt{d_i}|\qbi(j) - \halfqi(j)| \leq \gamma t$.
	\item If $w \geq (c\ln n/\gamma^2)m^{1+\mu}$, then we can get
	an estimate $\qbi(j)$ such that $\sqrt{d_j}|\qbi(j) - \qi(j)| \leq \beta_j$,
	where $\beta_j :=\sqrt{\frac{\gamma^2\max\{p_j, 1/m^{1+\mu}\}}{m^{1+\mu}}}$.
\end{itemize}
\end{claim}

\begin{myproof} We define a vector of random variables $X_k$, one
for each walk.
Define random variables $X_k(j)$ as follows:
$$X_k(j) = \begin{cases}
1 & \text{ walk $k$ ends at $j$ with even hops} \\
-1 & \text{ walk $k$ ends at $j$ with odd hops} \\
0 & \text{ walk $k$ doesn't end at $j$}
\end{cases}
$$
Note that $\av[X_k(j)] = \halfqi(j)d_j$, and $\var[X_k(j)] = p_j$. Our estimate
$\qbi(j)$ will be $\frac{1}{w}\sum_k X_k(j)$.
Observing that $|X_k(j)| \leq 1$, Bernstein's inequality implies that
for any $\beta > 0$,
%
$$\pr\left[\left|\frac{1}{wd_j}\sum_{k=1}^w X_k(j) - \halfqi(j)\right| > \beta \right]\
\leq\ 2\exp\left(-\frac{3w\beta^2d_j^2}{6p_i(j) + 2\beta d_j}\right). $$
%
For the first part, we set $\beta = \gamma t/\sqrt{d_j}$. For a sufficiently large $c$,
We get that the exponent is at least $4\ln n$, and hence the probability
is at most $1/n^4$. For the second part, we set $\beta = \beta_j/\sqrt{d_j}$.
Note that if $p_j < 1/m^{1+\mu}$, then $\beta_j < 1/m^{1+\mu}$.
So, the exponent is at least $4\ln n$, completing the proof.
\end{myproof}\\


We need to find a vector with a large Rayleigh
quotient that can be used in Lemma~\ref{lem:soto}. We already have a candidate
vector $\qi$. Although we get a very good approximation of this, note that
the order of vertices in an approximation can be very far from $\qi$.
Nonetheless, the following lemma allows us to do so.

\begin{claim} \label{clm:close}
Let $x$ be a vector such that $x^\top \lap x \geq (2 - \epsilon)\|x\|^2$. Then,
if $x'$ is a vector such that $\|x - x'\| < \delta\|x\|$, then
$\|x'\|^2 \geq (1-3\delta) \|x\|^2$ and $x'^\top \lap x' \geq (2 - \epsilon - 12\delta)\|x'\|^2$.
\end{claim}
\begin{myproof}
We have
$$x'^\top \lap x' - x^\top\lap x\ =\ x'^\top \lap x' - x'^\top \lap x + x'^\top \lap x - x^\top\lap x\\
\ =\ (x' - x)^\top \lap (x + x').$$ Thus,
$$|x'^\top \lap x' - x^\top\lap x|\ \leq\ (\|x'\| + \|x\|)\cdot \|\lap\| \cdot
\|x - x'\|\ \leq\ (2 + \delta)\|x\| \cdot 2 \cdot \delta\|x\|\ \leq\
6\delta\|x\|^2.$$ Furthermore,
$$|\|x'\|^2 - \|x\|^2|\ \leq\ (\|x'\| + \|x\|)\cdot \|x - x'\|\ \leq\ (2 + \delta)\|x\| \cdot \delta\|x\|\ \leq\
3\delta\|x\|^2.$$ Thus, we have
$$x'^\top \lap x'\ \geq\ x^\top \lap x - 6\delta \|x\|^2\ \geq\ (2-\epsilon -
6\delta)\|x\|^2\ \geq\ \frac{(2-\epsilon-6\delta)}{(1+3\delta)}\|x'\|^2\ \geq\ (2 - \epsilon - 12\delta)\|x'\|^2.$$
\end{myproof}

Now we prove Lemma~\ref{lem:good-thresh}.

\begin{myproof} Our cutting procedure is somewhat different from the
sweep cut used in~\cite{Tre09}. The most naive cut algorithm would
take $q_i$ and perform a sweep cut. Lemma~\ref{lem:good-q} combined
Lemma~\ref{lem:soto} would show that we can get a good cut. Unfortunately,
we are using an approximate version of $\halfqi$ ($\qbi$) for this purpose.
Nonetheless, Claim~\ref{clm:est-qi} tells us that we can get
good estimates of $\halfqi$, so $\qbi$ is close to $\halfqi$. Claim~\ref{clm:close}
tells us that $\qbi$ is good enough for all these arguments to go through
(since Lemma~\ref{lem:soto} only requires a bound on the Rayleigh quotient).

Our algorithm {\Find} is performing a geometric search for the right
threshold, invoking {\Threshold} many times. In each call
of the {\Threshold}, let estimate vector $\qbi^{(r)}$
be generated. Using these, we will construct a vector $\qti$.
This construction is not done by the algorithm, and is only
a thought experiment to help us analyze {\Find}.

Initially, all coordinates of $\qti$ are not defined, and we
incrementally set values. We will call {\Threshold}$(i,t_r)$
in order, just as {\Find}. In the call to {\Threshold}$(i,t_r)$,
we observe that vertices which are classified. These are
the vertices $j$ for which $\qbi^{(r)}(j) > t_r$ and
which have not been classified before. For all such $j$,
we set $\qti(j) := t_r$. We then proceed to the next
call of {\Threshold} and keep continuing until
the last call. After the last invocation of {\Threshold},
we simply set any unset $\qti(j)$ to $0$.

\begin{claim} \label{clm:qti} $\|\Dhalf \qti - \qi\| \leq 7\gamma \|\qi\|$
\end{claim}

\begin{myproof} Suppose $\halfqi(j) > t_r (1+4\gamma)$. Note that $\|\qbi^{(r-1)}(j) - \qi(j)\| \leq \gamma t_{r-1}/\sqrt{j}$.
Therefore,
$$\qbi^{(r-1)}(j) > t_r (1+4\gamma) - \gamma t_{r-1} > t_r(1+4\gamma) - \gamma (1+2\gamma) t_r
\geq t_r (1+2\gamma) \geq t_{r-1}$$
So $\qti(j)$ must be set in round $r-1$, if not before. If $\qti(j)$ remains
unset to the end (and is hence $0$), then $\halfqi(j) \leq t_r (1+4\gamma) $.
This implies that $\qi(j) \leq 2\gamma/m^{1+\mu/2}$. The total contribution
of all these coordinates to the difference $\|\Dhalf \qti - \qi\|^2$ is
at most $4\gamma^2/m^{1+\mu} \leq 4\gamma^2 \|\qi\|^2$.

Suppose $\qti(j)$ is set in round $r$ to $t_r$. This means
that $\qbi^{(r)}(j) > t_r$. By the choice of $w(t_r)$ and Claim~\ref{clm:est-qi},
$\sqrt{d_j}|\qbi^{(r)}(j) - \halfqi(j)| \leq \gamma t_r$.
Therefore,
\begin{eqnarray*}
	& |\sqrt{d_j} \qbi^{(r)}(j) - \qi(j)| \leq \gamma t_r \leq 2\gamma \qi(j) \\
	\Longrightarrow & \sqrt{d_j} \qbi^{(r)}(j) \leq (1+2\gamma) \qi(j) \\
	\Longrightarrow & \sqrt{d_j} \qti(j) = \sqrt{d_j} t_r \leq (1+2\gamma) \qi(j)
\end{eqnarray*}
Combining with the first part, we get $|\sqrt{d_j} \qti(j) - \qi(j)| \leq 5\gamma \qi(j)$.
\end{myproof}\\

We now observe that sweep cuts in $\qti$ generate \emph{exactly}
the same classifications that {\Threshold}$(i,t_r)$ outputs.
Therefore, it suffices to analyze sweep cuts of $\qti$.
We need to understand why there are thresholds that cut away many vertices.
Observe that the coordinates of $\qti$ are of the form $(1-\gamma)^r$.
This vector partitions all vertices in a natural way.
For each $r$, define $R_r := \{ j | \qti(j) = t_r\}$.
Call $r$ \emph{sparse}, if
$$(\sum_{j \in R_r} d_j) t_r^2 \leq \frac{\gamma^3}{m^{1+\mu}\log n}$$
Otherwise, it is \emph{dense}.
Note that a dense threshold exactly satisfies the condition
in Lemma~\ref{lem:good-thresh}.
Abusing notation, we call a vertex $j$ sparse if $j \in R_r$, such
that $r$ is sparse. Similarly, a threshold $t_r$ is sparse if $r$
is sparse. We construct a vector $\dense$.
If $j \in R_r$, for $r$ sparse, then $\dense(j) := 0$.
Otherwise, $\dense(j) := \qti(j)$.

\begin{claim} \label{clm:dense} $\|\Dhalf(\dense - \qti)\| \leq 2\gamma \|\qi\|$
\end{claim}

\begin{myproof}
\begin{eqnarray*} \|\Dhalf(\dense - \qti)\|^2 = \sum_{j : \dense(j)=0} d_j \qti(j)^2
& = & \sum_{r : r \ \text{sparse}} \sum_{j \in R_r} d_j \qti(j)^2)
= \Delta\sum_{r : r \ \text{sparse}} \sum_{j \in R_r} d_j t_r^2 \\
& \leq & \frac{4\log n}{\gamma} \cdot \frac{\gamma^3}{m^{1+\mu}\log n}
= \frac{4\gamma^2}{m^{1+\mu}\log n}
\leq \gamma^2 4\|\qi\|^2
\end{eqnarray*}
\end{myproof}

Let us now deal with the vector $\dense$ and perform the sweep cut
of~\cite{Tre09}.
All coordinates of $\dense$ are at most $1$. We choose a threshold
$t$ at random: we select $t^2$ uniformly at random\footnote{Both~\cite{Tre09}
and~\cite{Sot09} actually select $t$ uniformly at random, and use $\sqrt{t}$ as
a threshold. We do this modified version because it is more natural, for our
algorithm, to think of the threshold as a lower bound on the probabilities
we can detect.} from $[0,1]$. We do a rounding
to get the vector $z_t \in \{-1,0,1\}^n$:
$$z_t(j) = \begin{cases}
1 & \text{if} \  \qhi(j) \geq t \\
-1 & \text{if} \  \qhi(j) \leq -t \\
0 & \text{if} \  |\qhi(j)| < t
\end{cases}
$$
The non-zero vertices in $z_t$ are classified accordingly.
A \emph{cut} edge is one both of whose endpoints are non-zero
and of opposite size. A \emph{cross} edge is one where
only one endpoint is zero.
This classifying procedure is shown to cut a large
fraction of edges. By Lemma~\ref{lem:good-q}, we have
$\qi^\top \lap \qi \geq 2(1-\bar{\epsilon}) \|\qi\|^2$ (where $\bar{\epsilon}$
is some function of $\epsilon$ and $\mu$).
By Claims~\ref{clm:qti},~\ref{clm:dense} and Claim~\ref{clm:close},
$(\Dhalf\qhi)^\top \lap (\Dhalf\qhi) \geq 2(1-\bar{\epsilon}-c\gamma) \|\Dhalf\qhi\|^2$.
Then, by Lemma~\ref{lem:soto}, there are good thresholds for $\qhi$.
It remains to prove the following claim.

\begin{claim} \label{clm:couple} There are thresholds for $\qti$
that are dense \emph{and} good.
\end{claim}

\begin{myproof} We follow the analysis of~\cite{Sot09}.
We will perform sweep cuts for both $\qti$ and $\qhi$
and follow their behavior.
First, let take the sweep cut over $\qhi$.
Consider the indicator random variable $C(j,k)$ (resp. $X(j,k)$) that is $1$
if edge $(j,k)$ is a cut (resp. cross) edge. It is then
show that $\E[C(j,k) + \beta X(j,k)] \geq \beta(1-\beta)(\dense(j) - \dense(k))^2$,
where the expectation is over the choice of the threshold $t$.
Let us define a slight different choice of random thresholds.
As before $t^2$ is chosen uniformly at random from $[0,1]$.
Then, we find the smallest $t_r$ such that $r$ is dense
and $t_r \geq t$. We use this $t^* := t_r$ as the threshold for the cut.
Observe that this gives the \emph{same} distribution over
cuts as the original and \emph{only} selects dense thresholds. This is because in $\qhi$ all non-dense
vertices are set to $0$. All thresholds strictly in
between two consective dense $t_r$'s output the same
classification. The expectations of $C(j,k)$ and $X(j,k)$
are still the same.

We define analogous random variables $C'(j,k)$ and $X'(j,k)$ for $\qti$.
We still use the distribution over dense thresholds
as described above. When both $j$ and $k$ are dense,
we note that $C'(j,k) = C(j,k)$ and $X'(j,k) = X(j,k)$.
This is because if $t$ falls below, say, $\qti(j)$ (which is equal to
$\qbi(j)$), then $j$ will be cut. Even though $t^* > t$,
it will not cross $\qti(j)$, since $j$ is dense.
So, we have $\E[C'(j,k) + \beta X'(j,k)] = \E[C(j,k) + \beta X(j,k)]$.

If both $j$ and $k$ are not dense, then $C'(j,k) = X'(j,k) = 0$.
Therefore, $\E[C(j,k) + \beta X(j,k)] \geq \E[C'(j,k) + \beta X'(j,k)]$.
That leaves the main case, where $k$ is dense but $j$ is not.
Note that $\E[C(j,k)] = 0$, since $\qhi(j) = 0$.
We have $\E[X(j,k)] = \qhi(k)^2 = \qti(k)^2$.
If $|\qti(j)| \leq |\qti(k)|$, then
$\E[X'(j,k)] = \qti(k)^2 - \qti(j)^2$.
If $|\qti(j)| \leq |\qti(k)|$, then
$\E[X'(j,k)] \geq 0 \geq \qti(k)^2 - \qti(j)^2$.
So, we can bound $\E[X'(j,k)] \geq \E[X(j,k)] - \qti(j)^2$
and $\E[C'(j,k) + \beta X'(j,k)] \geq \beta(1-\beta)(\qti(j) - \qti(k))^2 - \beta \qti(j)^2$.

Summing over all edges, and applying the bound in Lemma 4.2 of~\ref{lem:soto}
for the non-prime random variables (dealing with $\qhi$), we get
\begin{eqnarray*}
\E[\sum_{(j,k)} C'(j,k) + \beta X'(j,k)]
& \geq & \E[\sum_{(j,k)} C(j,k) + \beta X(j,k)] - \beta \sum_{j \ \textrm{sparse}} d_j\qti(j)^2\\
& \geq & \beta (1-\beta) \sum_{(j,k) \ \textrm{edge}} (\qhi(j)-\qhi(k))^2 - \beta \gamma^2\|\Dhalf \qti\|^2 \\
& = & \beta (1-\beta) (\Dhalf\qhi)^\top \lap (\Dhalf\qhi) - \beta \gamma^2\|\Dhalf \qti\|^2 \\
& \geq & 2(1-\hat{\sigma}) \beta (1-\beta) \|\Dhalf \qhi\|^2 - 4\beta(1-\beta) \gamma^2 \|\Dhalf \qhi\|^2 \\
& \geq & 2(1-\sigma) \beta (1-\beta) \|\Dhalf \qti\|^2
\end{eqnarray*}
The second last step comes from the bound on $(\Dhalf\qhi)^\top \lap (\Dhalf\qhi)$ we have found,
and the observation that $\beta$ will always be set to less than $1/2$.
We have $1 - \hat{\sigma} = e^{-(2\delta+\mu)}(1-\epsilon) - O(\gamma)$ (based on Lemma~\ref{lem:good-q}.
Since $|\sigma - \hat{\sigma}| = O(\gamma)$, we get $\sigma$ as given
in Lemma~\ref{lem:find}. Because of the equations above, the analysis
of~\cite{Sot09} shows that the randomly chosen threshold $t^*$
has the property that
$$ \cut(P(\qti,t^*),N(\qti,t^*)) \geq \soto(\sigma)\inc(P(\qti,t^*),N(\qti,t^*))$$
Therefore, some threshold satisfies the condition 2(b) of {\Find}.
Note that the thresholds are chosen over a distribution of dense
thresholds. Hence, there is a good and dense threshold.
\end{myproof}

\end{myproof}

\section{\CutOrBound and local partitioning}
\def\ptil{\tilde{p}}
\def\Itil{\tilde{I}}
\def\Stil{\tilde{S}}
We describe our local partitioning procedure
{\CutOrBound} which is used to get the improved
running time. We first set some notation.
For a subset of vertices $S \subseteq V$, define $\bar{S} = V \setminus S$, and let $E(S, \bar{S})$ be the \emph{set} of edges crossing the cut $(S, \bar{S})$. Define the {\em weight} of $S$ to be $\volls(S) = 2\vol(S)$, to account for the self-loops of weight $1/2$: we assume that each vertex has a self-loop of weight
$d_i$, and the random walk simply chooses one edge with probability
proportional to its weight. For convenience, given a vertex $j$, $\volls(j) = \volls(\{j\}) = 2d_j$.
For a subset of edges $F \subseteq E$, let $\volls(F) = \sum_{e \in F} w_e$. The conductance of the set
$S$, $\phi_S$, is defined to be
$\phi_S = \frac{\volls(E(S, \bar{S}))}{\min\{\volls(S), \volls(\bar{S})\}}$.

\begin{center}
\fbox{\begin{minipage}{\columnwidth} \CutOrBound \ \ \ \ \ {\bf Input:} Graph $G$. \ \ \ \ {\bf Parameters:} Starting vertex $i$, $\alpha = m^{-\tau}, \ell = \ln(m)/\zeta$.
\begin{enumerate*}
    \item Define $\phi$ to satisfy $-\log(\frac{1}{2}(\sqrt{1-2\phi} + \sqrt{1+2\phi})) = \zeta\tau$, $w = \lceil 30\ell^2\ln(n)/\alpha \rceil= O(\log^3(n)/\alpha)$, $b = \lceil \frac{\ell}{2(1 - 2\phi)\alpha} \rceil = O(\log(n)/\alpha)$.

	\item Run $w $ random walks of length $\ell$ from $i$.

	\item For each length $l = 0, 1, 2, \ldots, \ell$:
	\begin{enumerate*}
		\item For any vertex $j$, let $w_j$ be the number of walks of length $l$ ending at $j$. Order the vertices in decreasing order of the ratio of $w_j/d_j$, breaking ties arbitrarily.

		\item For all $k \leq b$, compute the conductance of
		the set of top $k$ vertices in this order.

		\item If the conductance of any such set is less than $\phi$, stop and
		output the set.
	\end{enumerate*}

	\item Declare that $\max_j \frac{p_j}{2d_j} \leq 256\alpha.$
\end{enumerate*}
\end{minipage}}
\end{center}
The main theorem of this section is:
\begin{theorem} \label{thm:max}
Suppose a lazy random walk is run from a vertex $i$ for $\ell = \ln(m)/\zeta$ steps, for some constant $\zeta$. Let $p^\ell$ be the probability distribution induced on the final vertex.
Let $\alpha = m^{-\tau}$, for constant $\tau < 1$, be a given parameter so that $\zeta\tau < 1/8$, and let $\phi$ be chosen to satisfy $-\log(\frac{1}{2}(\sqrt{1-2\phi} + \sqrt{1+2\phi})) = \zeta\tau$.
Then, there is an algorithm \CutOrBound, that with probability $1 - o(1)$, in $O(\log^4(n)/\alpha)$ time, finds a cut
of conductance less than $\phi$, or declares correctly that $\max_j \frac{p^\ell_j}{2d_j} \leq 256\alpha$.
\end{theorem}
We provide a sketch before giving the detailed proof.
We use the Lov\'{a}sz-Simonovits curve technique~\cite{LS90}.
For every length $l = 0, 1, \ldots, \ell$, let $p^l$ be the probability vector induced on vertices after running a random walk of length $l$.
The Lov\'{a}sz-Simonovits curve $I^l:[0, 2m] \rightarrow
[0, 1]$ is constructed as follows. Let $j_1, j_2, \ldots, j_n$ be an ordering of the vertices
such that $\frac{p^l_{j_1}}{\volls(j_1)}\ \geq\ \frac{p^l_{j_2}}{\volls(j_2)}\
\geq\ \cdots\ \geq\ \frac{p^l_{j_1}}{\volls(j_n)}$.

For $k \in \{1, \ldots,
n\}$, define the set $S^l_k = \{j_1, j_2, \ldots, j_k\}$. For convenience, we
define $S^l_0 = \emptyset$, the empty set. For a subset of vertices $S$, and a
probability vector $p$, define $p(S) = \sum_{i \in S} p_i$. Then, we define
the curve $I^l$ at the following points: $I^l(\volls(S^l_k))\ :=\ p^l(S^l_k)$,
for $k = 0, 1, 2, \ldots, n.$ Now we complete the curve $I^l$ by
interpolating between these points using line segments.
Note that this curve is concave because the slopes of the line segments are
decreasing. Also, it is an increasing function. Lov\'{a}sz and Simonovits
prove that as $l$ increases, $I^l$ ``flattens" out, at a rate
governed by the conductance. A flatter $I^l$ means that the probabilities
at vertices are more equal (slopes are not very different), and hence
the walk is mixing.

Roughly speaking, the procedure {\CutOrBound} only looks the portion
of $I^l$ upto $S^l_b$, since it only tries to find sweep cuts among the
top $b$ vertices. We would like to argue that if {\CutOrBound}
is unsuccessful in finding a low conductance cut there, the
maximum probability should be small. In terms of the $I^l$s,
this means that the portion upto $S^l_b$ flattens out rapidly.
In some sense, we want to prove versions of theorems
in~\cite{LS90} that only talk about a prefix of the
$I^l$ curves.

The issue now is that it is not possible to compute the $p^l_j$'s (and $I^l$)
exactly since we only use random walks. We run walks
of length $l$ and get an empirical distribution
$\ptil^l$. We define $\Itil^l$ to be the
corresponding Lov\'{a}sz-Simonovits curve corresponding to
$\ptil^l$. If we run sufficiently many random walks and aggregate them to compute $\ptil^l_j$, then concentration bounds imply that $p^l_j$ is close
to $\ptil^l_j$ (when $p^l_j$ is large enough). Ideally, this should imply
that the behavior of $\Itil^l$ is similar to $I^l$.
There is a subtle difficulty here. The \emph{order}
of vertices with respect to $p^l$ and $\ptil^l$ could
be very different, and hence prefixes in the $I^l$ and
$\Itil^l$ could be dealing with different subsets
of vertices. Just because $I^l$ is flattening, it is
not obvious that $\Itil^l$ is doing the same.

Nonetheless, because for large $p^l_j$'s,
$\ptil^l_j$ is a good approximation, some sort
of flattening happens for $\Itil^l$. We give some
precise expressions to quantify this statement. Suppose
{\CutOrBound} is unable to find a cut of conductance $\phi$.
Then we show that for any $x \in [0, 2m]$, if $\hat{x} = \min\{x, 2m - x\}$,
$$\Itil^l(x)\ \leq\ \frac{e^{3\delta}}{2}(\Itil^{l-1}(x - 2\phi \hat{x}) + \Itil^{l-1}(x + 2\phi \hat{x})) + 4\delta\alpha x.$$
This is the flattening from $l-1$ to $l$. Since $\Itil^{l-1}$ is concave, the
averaging in the first part shows that $\Itil^l(x)$ is much smaller than
$\Itil^{l-1}(x)$. Note that additive error term, which does not occur
in~\cite{LS90}. This shows that when $x$ is large, this bound is not
interesting. That is no surprise, because we can only sample some prefix of
$I^l$. Then, we prove by induction on $l$ that, if we define $\psi =
-\log(\frac{1}{2}(\sqrt{1-2\phi} + \sqrt{1+2\phi})) = \zeta\tau$, then
$\Itil^l(x)\ \leq\ e^{3\delta l}\left[\sqrt{x}e^{-\psi l} + \frac{x}{2m}\right] + 4e^{4\delta l}\alpha x$.
Assuming that $\delta \approx 1/\ell$,  the $e^{-\psi l}$ term decays very
rapidly. For the final $\ell = \Omega(\log(n)/\psi)$, we are only left with the error term, which will be $O(\alpha)$. We then get
$\max_j \frac{\ptil^\ell_j}{2d_j}\ =\ \Itil^\ell(1)\ \leq\ O(e^{-\psi \ell} + \frac{1}{m} + \alpha)\ \leq\ O(\alpha)$.

\subsection{Proof of Theorem~\ref{thm:max}} \label{sec:ls}

First, we note that $\phi \leq \sqrt{2\zeta\tau}$, so $1 - 2\phi > 0$. Consider the following algorithm:

It is easy to see that this algorithm can be implemented to run in time $O(\log^4(n)/\alpha)$. We now prove that this algorithm has the claimed behavior. We make use of the Lov\'{a}sz-Simonovits curve technique. For every length $l = 0, 1, \ldots, \ell$, let $p^l$ be the probability vector induced on vertices after running a random walk of length $l$.

Now, we construct the Lov\'{a}sz-Simonovits curve~\cite{LS90}, $I^l:[0, 2m] \rightarrow
[0, 1]$ as follows. Let $j_1, j_2, \ldots, j_n$ be an ordering of the vertices
as follows:
$$\frac{p^l_{j_1}}{\volls(j_1)}\ \geq\ \frac{p^l_{j_2}}{\volls(j_2)}\
\geq\ \cdots\ \geq\ \frac{p^l_{j_1}}{\volls(j_n)}.$$
For $k \in \{1, \ldots,
n\}$, define the set $S^l_k = \{j_1, j_2, \ldots, j_k\}$. For convenience, we
define $S^l_0 = \emptyset$, the empty set. For a subset of vertices $S$, and a
probability vector $p$, define $p(S) = \sum_{i \in S} p_i$. Then, we define
the curve $I^l$ at the following points: $I^l(\volls(S^l_k))\ :=\ p^l(S^l_k)$,
for $k = 0, 1, 2, \ldots, n.$ Now we complete the curve $I^l$ by
interpolating between these points using line segments. Note that the slope of
the line segment of the curve at the points $\volls(S^l_k), \volls(S^l_{k+1})$ is
exactly $\frac{p^l_{j_{k+1}}}{\volls(j_{k+1})}$.  A direct definition of the
curve is the following: for any point $x \in [0, 2m]$, if $k$ is the unique
index where $x \in [\volls(S^l_k), \volls(S^l_{k+1}))$, then $I^l(x) = p^l(S^l_k) +
(x - \volls(S^l_k))\cdot \frac{p^l_{j_{k+1}}}{\volls(j_{k+1})}.$

An useful alternative definition for $I^l(x)$ is the following:
\begin{equation}\label{eq:alt-def}
I^l(x)\ =\ \max \sum_i p^l_i w_i \quad \text{subject to} \quad w_1, w_2, \ldots, w_n \in [0, 1];\ \sum_i \volls(i) w_i \leq x.
\end{equation}

Note that this curve is concave because the slopes of the line segments are
decreasing. Also, it is an increasing function. Now, Lov\'{a}sz and Simonovits prove the following facts about the
curve: let $S \subseteq V$ be any set of vertices, and let $x_S = \volls(S)$ and $\phi_S$ be its conductance.
For $x \in [0, 2m]$, define $\hat{x} =
\min\{x, 2m - x\}$. Then, we have the following:
\begin{equation} \label{eq:ls-chords}
p^l(S)\ \leq\ \frac{1}{2}(I^{l-1}(x_S - 2\phi_S \hat{x}_S) + I^{l-1}(x_S + 2\phi_S \hat{x}_S)).
\end{equation}
Furthermore, for any $x \in [0, 2m]$, we have
$I^l(x)\ \leq\ I^{l-1}(x)$.

The issue now is that it is not possible to compute the $p^l_j$'s exactly since we only use random walks. Fix an error parameter $\delta = 1/\ell$. In the algorithm \CutOrBound, we run $w = c\cdot \frac{1}{\alpha} \cdot \ln(n)$ walks of length $\ell$, where $c = 30/\delta^2$. For each length
$l$, $0 \leq l \leq \ell$, consider the empirical distribution $\ptil^l$
induced by the walks on the vertices of the graph, i.e. $\ptil^l_j = w_j/w$, where $w_j$ is the number of walks of length $l$ ending at $j$. We search for low conductance cuts by ordering the vertices in decreasing order of $\ptil^l$ and checking the sets of top $k$ vertices in this order, for all $k = 1, 2, \ldots, O(1/\delta\alpha)$. This takes time $O(w \ell)$. To show that this works,
first, define $\Itil^l$ be the Lov\'{a}sz-Simonovits curve corresponding to
$\ptil^l$. Then, we have the following:
\begin{Lem} \label{lem:concentration-ptil}
With probability $1 - o(1)$, the following
holds. For every vertex subset of vertices $S \subseteq V$, we have
$$(1-\delta)p^l(S) - \delta\alpha\volls(S)\ \leq\ \ptil^l_j\ \leq\ (1+\delta)p^l(S) + \delta\alpha\volls(S).$$
For every length $l$, and every $x \in [0, 2m]$,
$$(1-\delta)I^l(x) - \delta\alpha x\ \leq\ \tilde{I}^l(x)\ \leq\ (1+\delta)I^l(x) + \delta\alpha x.$$
\end{Lem}
\begin{myproof}
For any vertex $j$, define $\delta_j = \delta(p^l_j + \alpha)$. By Bernstein's
inequality, we have $$\pr[|\ptil^l_j - p^l_j| > \delta_j]\ \leq\
2\exp\left(-\frac{\delta_j^2w}{2p^l_j + 2\delta_j/3}\right)\ <\
2\exp(-\delta^2c\ln(n)/3) \leq 1/n^{10}$$ since $c = 30/\delta^2$. So
with probability at least $1 - o(1)$, for all lengths $l$, and for all vertices
$j$, we have $$(1-\delta)p^l_j - \delta\alpha\ \leq\ \ptil^l_j\ \leq\
(1+\delta)p^l_j + \delta\alpha.$$ Assume this is the case. This immediately
implies that for any set $S$, we have $$(1-\delta)p^l(S) - \delta\alpha|S|\
\leq\ \ptil^l(S)\ \leq\ (1+\delta)p^l(S) + \delta\alpha|S|.$$

Now, because both curves $I^l$ and $\Itil^l$ are piecewise linear, concave and
increasing, to prove the lower bound in the claimed inequality, it suffices to
prove it for only $x = x_k = \volls(S^l_k)$, for $k = 0, 1, \ldots, n$. So fix
such an index $k$.

Now, $I^l(x_k) = p^l(S^l_k)$. Consider $\ptil^l(S^l_k)$. We have
$$\ptil^l(S^l_k)\ \geq\ (1-\delta)p^l(S^l_k) - \delta\alpha|S^l_k|\ \geq\
(1-\delta)p^l(S^l_k) - \delta\alpha\volls(S^l_k).$$ Now, the alternative
definition of the Lov\'{a}sz-Simonovits curve (\ref{eq:alt-def}) implies that
$\Itil^l(\volls(S^l_k)) \geq \ptil^l(S^l_k)$, so we get $$\Itil^l(x_k)\ \geq\
(1-\delta)p^l(S^l_k) - \delta\alpha x_k,$$ as required. The upper bound is
proved similarly, considering instead the corresponding sets $\tilde{S}^l_k$
for $\Itil^l$ consisting of the top $k$ vertices in $\ptil^l$ probability.
\end{myproof}

The algorithm \CutOrBound can be seen to be searching for low conductance cuts in the top $b$ vertices in the order given by $\ptil^l_j/\volls(j)$. Now, we prove that if we only find large conductance cuts, then the curve $\Itil^l$ ``flattens'' out rapidly. Let $j'_1, j'_2, \ldots, j'_n$ be this order.
Let $\Stil^l_k = \{j'_1, j'_2, \ldots, j'_k\}$ be the set of top $k$ vertices in the order, $x_k = \volls(\Stil^l_k)$, and $\phi_k$ be the conductance of $\Stil^l_k$. Now we are ready to show our flattening lemma:
\begin{Lem} \label{lem:ls-chords-empirical} With probability $1 - o(1)$, the
following holds. Suppose the algorithm \CutOrBound finds only cuts of conductance $\phi$ when sweeping over the top $b$ vertices in $\ptil^l$ probability. Then, for any index $k = 0, 1, \ldots, n$, we have
$$p^l(\Stil^l_k)\ \leq\ \frac{1}{2}(I^{l-1}(x_k - 2\phi\hat{x}_k) + I^{l-1}(x_k + 2\phi\hat{x}_k)) + \delta\alpha \phi\hat{x}_k.$$
\end{Lem}
\begin{myproof}
Let $G = \left\{j:\ \frac{p^{l-1}_j}{\volls(j)} > \delta\alpha\right\}$.
We have $1 \geq p^{l-1}(G) > \delta\alpha\volls(G)$, so $\volls(G) < 1/\delta\alpha$.

As defined in the algorithm \CutOrBound, let $b = \lceil \frac{1}{2(1 - 2\phi)\delta\alpha}\rceil$. Let $a$ be the largest index so that $\ptil^l_{j'_{a}} > 0$. If $a < b$, then let $Z$ be the set of $b - a$ vertices $k$ of zero $\ptil^l$ probability considered by algorithm \CutOrBound for searching for low conductance cuts. We assume that in choosing the ordering of vertices to construct $\Itil^l$, the vertices in $Z$ appear right after the vertex $j'_{a}$. This doesn't change the curve $\Itil^l$ since the zero $\ptil^l$ probability vertices may be arbitrarily ordered.


Suppose that the algorithm \CutOrBound finds only cuts of conductance at least
$\phi$ when running over the top $b$ vertices. Then, let $k$ be some index in $0,
1, \ldots, n$. We consider two cases for the index
$k$:\\
{\bf Case 1:} $k \leq b$:\\
In this case, since the sweep only yielded cuts of conductance at least $\phi$, we have $\phi_k \geq \phi$. Then (\ref{eq:ls-chords}) implies that
$$p^l(\Stil^l_k)\ \leq\ \frac{1}{2}(I^{l-1}(x_k - 2\phi \hat{x}_k) + I^{l-1}(x_k + 2\phi \hat{x}_k)).$$

\noindent{\bf Case 2:}  $k > b$:\\
We have
$$x_k\ >\ x_{b}\ =\ \volls(\Stil^l_b)\ \geq\ 2b\ \geq \frac{1}{(1-2\phi)\delta\alpha}\ >\ \frac{1}{1-2\phi}\volls(G).$$ Thus,
$\volls(G) < (1- 2\phi)x_k\ \leq\ x_k - 2\phi\hat{x}_k$.
Hence, the slope of the curve $I^{l-1}$ at the point $x_k - 2\phi \hat{x}_k$
is at most $\delta\alpha$. Since the curve $I^{l-1}$ is concave and
increasing, we conclude that
$$I^{l-1}(x_k - 2\phi\hat{x}_k)\ \geq\ I^{l-1}(x_k) - 2\delta\alpha\phi\hat{x}_k,$$
and
$$I^{l-1}(x_k + 2\phi \hat{x}_k)\ \geq\ I^{t-1}(x_k).$$
Since $p^l(\Stil^l_k) \leq I^l(x_k) \leq I^{l-1}(x_k)$,
$$p^l(\Stil^l_k)\ \leq\ \frac{1}{2}(I^{l-1}(x_k - 2\phi\hat{x}_k) + I^{l-1}(x_k + 2\phi\hat{x}_k)) + \delta\alpha \phi \hat{x}_k.$$
This completes the proof of the lemma.
\end{myproof}

Since the bounds of Lemma~\ref{lem:concentration-ptil} hold with probability $1 - o(1)$, we assume from now on that is indeed the case for all lengths $l$. Thus, we conclude that if we never find a cut of conductance at most $\phi$, and for any index $k = 0, 1, \ldots, \ell$, we have
\begin{align*}
\Itil^l_k(x_k)\ &=\ \ptil^l_k(\Stil^l_k)\\
&\leq\ (1+\delta)p^l_k(\Stil^l_k) + \delta\alpha x_k & \text{by Lemma~\ref{lem:concentration-ptil}}\\
&\leq\ \frac{1+\delta}{2}(I^{l-1}(x_k - 2\phi\hat{x}_k) + I^{l-1}(x_k + 2\phi\hat{x}_k)) + 2\delta\alpha x_k & \text{by Lemma~\ref{lem:ls-chords-empirical}}\\
&\leq\ \frac{1+\delta}{2(1-\delta)}(\Itil^{l-1}(x_k - 2\phi\hat{x}_k) + \Itil^{l-1}(x_k + 2\phi\hat{x}_k)) + 4\delta\alpha x_k & \text{by Lemma~\ref{lem:concentration-ptil}}
\end{align*}
Here, we use the facts that $(1+\delta)\phi \leq 1$, and $\frac{1+\delta}{1-\delta} \leq 2$. Now, because $\Itil^l$ is a piecewise linear and concave
function, where the slope only changes at the $x_k$ points, the above
inequality implies that for all $x \in [0, 2m]$, we have
$$\Itil^l(x)\ \leq\ \frac{e^{3\delta}}{2}(\Itil^{l-1}(x - 2\phi \hat{x}) + \Itil^{l-1}(x + 2\phi \hat{x})) + 4\delta\alpha x.$$
Here, we used the bound $\frac{1+\delta}{1-\delta} \leq e^3\delta$.

Now, assume that we never find a cut of conductance at most $\phi$ over all lengths $l$. Define $\psi = -\log(\frac{1}{2}(\sqrt{1-2\phi} + \sqrt{1+2\phi})) = \zeta\tau$. Note that $\psi \geq \phi^2/2$. Then, we prove by induction on $l$ that
$$\Itil^l(x)\ \leq\ e^{3\delta l}\left[\sqrt{\hat{x}}e^{-\psi l} + \frac{x}{2m}\right] + 4e^{4\delta l}\alpha x.$$ The statement for $l = 0$ is easy to see, since the curve
$I^0(x) = \min\{x/2d_i, 1\}$ (recall that we start the walk at vertex $i$).
Assuming the truth of this bound for $l-1$, we now show it for $l$. We have
\begin{align*}
\Itil^l(x)\ &\leq\ \frac{e^{3\delta}}{2}(\Itil^{l-1}(x - 2\phi \hat{x}) + \Itil^{l-1}(x + 2\phi \hat{x})) + 4\delta\alpha x \\
&\leq\ \frac{e^{3\delta}}{2}\left[e^{3\delta(l-1)}\left[\sqrt{\widehat{(x - 2\phi\hat{x})}}e^{-\psi(l-1)} + \sqrt{\widehat{(x + 2\phi\hat{x})}}e^{-\psi(l-1)}
+ \frac{2x}{2m}\right] + 8e^{4\delta(l-1)}\alpha x\right]\\
&\quad + 4\delta\alpha x \\
&\leq\ e^{3\delta l}\left[\sqrt{\hat{x}}e^{-\psi l} + \frac{x}{2m}\right] + 4 e^{4\delta l}\alpha x,
\end{align*}
which completes the induction. In the last step, we used the following bounds:
if $x \leq m$, then
$$\sqrt{\widehat{(x - 2\phi\hat{x})}} + \sqrt{\widehat{(x +2\phi\hat{x})}}\
\leq\ \sqrt{x - 2\phi x} + \sqrt{x +2\phi x}\ =\ 2\hat{x}e^{-\psi},$$ and if $x > m$, then
$$\sqrt{\widehat{(x - 2\phi\hat{x})}} + \sqrt{\widehat{(x +2\phi\hat{x})}}\
\leq\ \sqrt{2m - (x - 2\phi(2m - x))} + \sqrt{2m - (x + 2\phi(2m - x))}\
=\ 2\hat{x} e^{-\psi}.$$
Since $\delta = 1/\ell$, we get
$$\max_j \frac{\ptil^\ell_j}{2d_j}\ =\ \Itil^\ell(1)\ \leq\ e^{-\psi \ell + 3} + \frac{e^3}{2m} + 4e^4\alpha\ \leq\ 250\alpha,$$
assuming $\alpha = m^{-\tau}$, $\ell = \frac{\ln m}{\zeta}$, and $\psi = \zeta\tau$. Finally, again invoking Lemma~\ref{lem:concentration-ptil}, we get that $\max p^l_j/2d_j \leq 256\alpha$, since $\delta = 1/\ell$.

\section{Recursive partitioning} \label{sec:algo}

Given the procedure {\Find}, one can construct
a recursive partitioning algorithm to approximate the
{\maxcut}.
We classify some vertices through {\Find}, remove them,
and recurse on the rest of the graph. We call
this algorithm {\Simple}. The algorithm {\Balance}
uses the low conductance sets obtained from
Theorem~\ref{thm:max} and does a careful
balancing of parameters to get an improved
running time. All proofs of this section, including theoretical guarantees on approximation factors, are in Appendix~\ref{app:algo}. We state the procedure {\Simple}
first and provide the relevant claims.

\begin{center}
\fbox{\begin{minipage}{\columnwidth} \Simple \ \ \ \ \ {\bf Input:} Graph $G$. \ \ \ \ {\bf Parameters:} $\epsilon, \mu, \alpha$.
\begin{enumerate*}
		\item If $\soto(\sigma(\epsilon,\mu)) = 1/2$, then put each vertex
		in $L$ or $R$ uniformly at random (and return).
    \item Let $P$ be a set of $O(\log n)$ vertices chosen uniformly at random.
    \begin{enumerate*}
    	\item For all $i \in P$, run procedures {\Find}$(i,\mu)$ in parallel. Stop when any one of these
    	succeeds or all of them fail.
    \end{enumerate*}
    \item If all procedures failed, output FAIL.
    \item \label{step} Let the successful output be the set $Even_i$ and $Odd_i$. With
    probability $1/2$, put $Even_{i}$ in $L$ and $Odd_{i}$ in $R$. With probability $1/2$,
    do the opposite.
    \item Let $\xi = 1-\inc(Even_i,Odd_i)/m$. Set $\epsilon' = \epsilon/\xi$ and $G'$ be the induced
    subgraph on unclassified vertices. Run {\Simple}$(G',\epsilon',\mu)$.
    If it succeeds, output the final cut $L$ and $R$.
    \item If $G$ is the original graph, put each vertex (even those already classified)
    randomly in $L$ or $R$. Irrespective of $G$, output FAIL.
\end{enumerate*}
\end{minipage}}
\end{center}

The guarantees
of {\Simple} are in terms of a function $H(\epsilon,\mu)$.
For a given $\epsilon$ and $\mu$, let $z^*$ be the largest value
such that $\soto(\sigma(\epsilon/z^*,\mu)) = 1/2$.
Then $H(\epsilon,\mu) := z^*/2 + \int_{z^*}^1 \soto(\sigma(\epsilon/z,\mu)) dz$. For constant $\epsilon < 0.5$, $H(\epsilon,\mu)$ is a constant
$> 0.5$.

\begin{Lem} \label{lem:simple} Let $\maxcut(G) = 1-\epsilon$.
There is an algorithm {\Simple}$'(G,\mu)$ that, with high probability,
outputs a cut of value $H(\epsilon,\mu) - o(1)$, and thus the worst-case approximation ratio is $\min_\epsilon \frac{H(\epsilon, \mu)}{1-\epsilon} - o(1)$. The running time is $\otilde(\Delta m^{2+\mu})$.
\end{Lem}
Tle algorithm {\Simple}$'$ is a version of {\Simple} that only takes $\mu$ as a
parameter and searches for the appropriate value of $\epsilon$. Suppose
$\maxcut(G)=1-\epsilon$. The procedure {\Simple}$'$ runs
{\Simple}$(G,\epsilon_r,\mu, 1)$ (i.e. $\alpha = 1$), for all $\epsilon_r$ such
that $1-\epsilon_r = (1-\gamma)^r$ and $1/2 \leq  1-\epsilon_r \leq 1$. By
choosing $\gamma$ small enough and Claim~\ref{clm:simple} below, we can ensure
that we cut at least $H(\epsilon, \mu) - o(1)$ fraction of edges. It therefore
suffices to prove:
\begin{claim} \label{clm:simple}
If {\Simple}$(G,\epsilon,\mu)$ succeeds, it outputs a cut of (fractional) value at least
$H(\epsilon,\mu)$. If it fails, it outputs a cut of value $1/2$. If
$\maxcut(G)$ $\geq 1-\epsilon$, then {\Simple}$(G,\epsilon,\mu)$ succeeds with
high probability. The running time is always bounded by $\otilde(\Delta
m^{2+\mu})$.
\end{claim}
%

We now describe {\Balance} and state the main lemma associated
with it. We observe that {\Balance} uses \CutOrBound to
either decompose the graph into pieces, or ensure
that we classify many vertices. We use Theorem~\ref{thm:max}
to bound the running time.

\begin{center}
\fbox{\begin{minipage}{\columnwidth} \Balance \ \ \ \ \  {\bf Input:} Graph $G$.  \ \ \ \
{\bf Parameters:} $\epsilon_1,\mu_1,\epsilon_2,\mu_2,\alpha = m^{-\tau}$.
\begin{enumerate*}
    \item Let $P$ be a random subset of $O(\log n)$ vertices.
    \item For each vertex $i \in P$, run {\CutOrBound}$(i,\ell(\epsilon_1,\mu_1),\alpha)$.
    \item If a low conductance set $S$ was found by any of the above calls:
    \begin{enumerate*}
    		\item Let $G_S$ be the induced graph on $S$, and $G'$ be the induced graph on $V\setminus S$.
    Run {\Simple}$'(G_S,\mu_2)$ and {\Balance}$(G')$ (with same parameters) to get the final partition.
    \end{enumerate*}
   	\item Run {\Simple}$(G,\epsilon_1,\mu_1, \alpha)$ up to Step~\ref{step}, using random vertex set $P$.
   	Then run \Balance$(G')$ (with same parameters), where
		$G'$ is the induced graph on the unclassified vertices.
		\item Output the better of this cut and the trivial cut.
\end{enumerate*}
\end{minipage}}
\end{center}

\begin{Lem} \label{lem:balance}
For any constant $b > 1.5$, there is a choice of $\mu_1$, $\mu_2$ and
$\tau$ so that {\Balance} runs in $\otilde(\Delta m^b)$ time and provides
an approximation factor that is a constant greater than $0.5$.
\end{Lem}

Let us give a simple explanation for the $1.5$-factor. Neglecting the $\mu$'s and
polylogarithmic factors, we perform $O(1/\alpha)$ walks in {\CutOrBound}. In
the worst case, we could get a low conductance set of constant size, in which
case the work per output is $O(1/\alpha)$. When we have the $\alpha$ bound on
probabilities, the work per output is $O(\alpha m)$. So it appears that $\alpha
= 1/\sqrt{m}$ is the balancing point, which yields an $\otilde(m^{1.5})$ time
algorithm.

In the next subsection, we define many parameters
which will be central to our analysis.
We then provide detailed proofs for Claim~\ref{clm:simple} and
Lemma~\ref{lem:balance}. Finally, we give a graph
detailing how the approximation factor increases with
running time (for both {\Simple} and {\Balance}).


\subsection{Preliminaries} \label{app:algo}

For convenience, we list the various free parameters and dependent variables.
\begin{itemize*}
	\item $\epsilon$ is the maxcut parameter, as described above. Eventually, this will be set
	to some constant (this is explained in more detail later).
	\item $\mu$ is a running time parameter. This is used to control the norm
	of the $\qti$ vector, and through that, the running time. This affects the approximation
	factor obtained, through Lemma~\ref{lem:good-q}.
	\item $\alpha (= m^{-\tau})$ is the maximum probability parameter. This directly affects the running
	time through Lemma~\ref{lem:find}. For {\Simple}, this is just set to $1$, so it
	only plays a role in {\Balance}.
	\item $\ell(\epsilon,\mu) := \mu(\ln(4m/\delta^2)/[2(\delta+\epsilon)]$. This is the
	length of the random walk.
	\item $\sigma(\epsilon,\mu)$ is the parameter that is in Lemma~\ref{lem:find}. Setting
	$\epsilon' = -\ln(1-\epsilon)/\mu$, we get $ 1-\sigma = e^{-\epsilon'}(1-\epsilon)(1-\delta)(1-\gamma) $.
	\item $\chi(\epsilon,\mu,\alpha)$ is the cut parameter that comes from Theorem~\ref{thm:max}.
	When we get a set $S$ of low conductance, the number of edges in the cut
	is at most $\chi(\epsilon,\mu)|Internal(S)|$. Here, $Internal(S)$ is the set
	of edges internal to $S$. In Theorem~\ref{thm:max}, the number
	of cut edges in stated in terms of the conductance $\phi$. We have
	$\chi = 4\phi/(1-2\phi)$. Also, $\phi$ is at most $\sqrt{4\epsilon\tau/\mu}$.
	We will drop the dependence on $\alpha$, since it will be fixed (more details
	given later).
\end{itemize*}

We will also use some properties of the function $H(\epsilon,\mu)$.

\begin{Lem} \label{lem:H-properties}
For any fixed $\mu > 0$, $H(\epsilon, \mu)$ is a convex, decreasing function of
$\epsilon$. Furthermore, there is a value $\bar{\epsilon} =
\bar{\epsilon}(\mu)$ such that $H(\bar{\epsilon}, \mu) > 0.5029$.
\end{Lem}
\begin{myproof}
First, note that $\soto(\sigma)$ is a decreasing function of $\sigma$. This is
because all the three functions that define $\soto$ are decreasing in their
respective ranges, and the transition from one function to the next occurs
precisely at the point where the functions are equal.

Now, for any fixed $\mu$, $\sigma(\epsilon, \mu)$ is a strictly increasing
function of $\epsilon$, and hence, $\soto(\sigma(\epsilon, \mu))$ is a
decreasing function of $\epsilon$. Thus, $H(\epsilon, \mu) = \int_0^1
\soto(\sigma(\epsilon/r, \mu))dr$ is a decreasing function of $\epsilon$, since
for any fixed $r$, the integrand $\soto(\sigma(\epsilon/r, \mu))$ is a
decreasing function of $\epsilon$.

For convenience of notation, we will use $H$ and $\sigma$ to refer $H(\epsilon,
\mu)$ and $\sigma(\epsilon, \mu)$ respectively. Now define $x = \epsilon/r$.
Doing this change of variables in the integral, we get $H = \epsilon
\int_\epsilon^\infty \frac{\soto(\sigma(x, \mu))}{x^2} dx$. By the fundamental
theorem of calculus, we get that
$$\frac{\partial H}{\partial \epsilon}\ =\ \int_\epsilon^\infty \frac{\soto(\sigma(x, \mu))}{x^2} dx - \frac{\soto(\sigma)}{\epsilon}.$$
Again applying the fundamental theorem of calculus, we get that
$$\frac{\partial^2 H}{\partial \epsilon^2}\ =\
-\frac{\soto(\sigma)}{\epsilon^2} - \frac{\epsilon\frac{\partial
\soto(\sigma)}{\partial \epsilon} - \soto(\sigma)}{\epsilon^2}\ =\
-\frac{1}{\epsilon}\cdot\frac{\partial \soto(\sigma)}{\partial \epsilon}\ \geq\
0,$$ since $\soto(\sigma)$ is a decreasing function of $\epsilon$. Thus, $H$ is a convex
function of $\epsilon$.

To show the last part, let $\sigma^{-1}_\mu$ is the inverse function of
$\sigma(\epsilon, \mu)$, keeping $\mu$ fixed, and consider $\bar{\epsilon}(\mu)
= \sigma^{-1}_{\mu}(1/4) = 1 - (\frac{3}{4})^{\frac{\mu}{1+\mu}} - o(1)$, by
making $\delta$ and $\gamma$ small enough constants. For $r \in [1/4, 1/3]$, we
have $\soto(\sigma(\bar{\epsilon}/r, \mu)) \geq \soto(1/4)
> 0.535$. Thus, we get
$$H(\bar{\epsilon}, \mu)\ >\ 0.5 + 0.035 \times (1/3 - 1/4)\ =\ 0.5029.$$
\end{myproof}

\subsection{Proof for {\Simple}}

As we showed in the main body, it suffices to prove
Claim~\ref{clm:simple}.

\begin{myproof} (of Claim~\ref{clm:simple}) This closely follows the analysis given
in~\cite{Tre09} and~\cite{Sot09}.
If any recursive call to \Simple fails, then the top
level algorithm also fails and outputs the trivial cut.

Suppose $\maxcut(G)$ is at least $1-\epsilon$.
Then $\maxcut(G')$ is at least
$$\frac{(1-\epsilon)m-\inc(Even_i,Odd_i)}{m-\inc(Even_i,Odd_i)}
= 1 - \epsilon/\xi $$
Applying this inductively, we can argue that \emph{whenever} a recursive call
{\Simple}$(G',\epsilon',\mu)$ is made, $\maxcut(G')$ $\geq 1-\epsilon'$.
From Lemma~\ref{lem:good-q}, since $O(\log n)$ vertices are chosen in $P$,
with high probability, in \emph{every} recursive call, a good vertex
is present in $P$. From Lemma~\ref{lem:find}, in every recursive call,
with high probability, some call to {\Find} succeeds. Hence, {\Simple} will
not output FAIL and succeeds.

Assuming the success of {\Simple}, let us compute the total number of edges cut. We denote the parameters of the $t$th
recursive call to {\Simple} by subscripts of $t$. Let the number of edges
in $G_t$ be $\rho_t m$ (where $\rho_0 = 1$). Let $T$ be the last call to {\Simple}.
We have $\epsilon_t = \epsilon/\rho_t$.
Only for $t = T$, we have that $\soto(\sigma(\epsilon/\rho_t,\mu)) = 1/2$.
In the last round, we cut
$\rho_T m/2$ edges.
The number of cut
edges in other rounds is $\soto(\sigma(\epsilon_t,\mu)) (\rho_t - \rho_{t+1}) m$.
Summing over all $t$, the total number of edges cut (as a fraction of $m$) is
\begin{eqnarray*}
	\sum_{t=0}^{T-1} \soto(\sigma(\epsilon_t,\mu)) (\rho_t - \rho_{t+1}) + \rho_T/2 & = &
	\sum_{t=0}^{T-2} \int_{\rho_{t+1}}^{\rho_t} \soto(\sigma(\epsilon/\rho_t,\mu)) dr +
	\int_{\rho_T}^{\rho_{T-1}} \soto(\sigma(\epsilon/\rho_t,\mu)) dr + \rho_T/2 \\
	& = & \sum_{t=0}^{T-2} \int_{\rho_{t+1}}^{\rho_t} \soto(\sigma(\epsilon/\rho_t,\mu)) dr +
	\int_{z_\mu^*}^{\rho_{T-1}} \soto(\sigma(\epsilon/\rho_t,\mu)) dr \\
	& & \int_{\rho_T}^{z_\mu^*} (1/2) dr + \rho_T/2 \\
	& \geq & \sum_{t=0}^{T-2} \int_{\rho_{t+1}}^{\rho_t} \soto(\sigma(\epsilon/r,\mu)) dr +
	\int_{z_\mu^*}^{\rho_{T-1}} \soto(\sigma(\epsilon/r,\mu)) dr + z_\mu^*/2 \\
	& = & \int_{z_\mu^*}^{1} \soto(\sigma(\epsilon/r,\mu)) dr	+ z_\mu^*/2
\end{eqnarray*}
The inequality comes about because $\soto$ is a decreasing function
and $\sigma$ is an increasing function of $\epsilon$.

We now bound the running time, using Lemma~\ref{lem:find}. Consider a successful iteration $t$.
Suppose the number of vertices classified in this iteration is $N_t$.
The total running time in iteration $t$ is
$\otilde(N_t\Delta m^{1+\mu})$. This is because we run the $O(\log n)$ calls
in parallel, so the running time is at most $O(\log n)$ times
the running time of the successful call. Summed over all
iterations, this is at most $\otilde(\Delta m^{2+\mu})$. Suppose
an iteration is unsuccessful, the total running time
is $\otilde(\Delta m^{2+\mu})$. There can only be one such iteration,
and the claimed bound follows.
\end{myproof}

\subsection{Proofs for {\Balance}}

We first give a rather complicated expression for the approximation ratio
of {\Balance}. First, for any $\mu > 0$, define $h(\mu) = \min_\epsilon \frac{H(\epsilon, \mu)}{1-\epsilon}$. This is essentially the approximation factor of \Simple'.

\begin{claim} \label{clm:threehalf} The algorithm {\Balance} has a work to output ratio
of $\otilde(\Delta (m^{\tau+\mu_2\tau} + m^{1+\mu_1-\tau}))$. The approximation ratio
is at least:
$$ \max_{\epsilon_1}\min\left\{
\min_{\epsilon} \max\left\{\frac{1}{2(1-\epsilon)},
\frac{h(\mu_2) (1-\epsilon-\epsilon\chi(\epsilon_1,\mu_1)) + \chi(\epsilon_1, \mu_1)/2}{(1-\epsilon)(1+\chi(\epsilon_1,\mu_1))}
\right\},
H(\epsilon_1,\mu_1), \frac{1}{2(1-\epsilon_1)}
\right\} $$
\end{claim}

\begin{myproof} First let us analyze the work per output ratio of \Balance.
We initially perform $\otilde(\Delta m^\tau)$ walks. Suppose we get a low
conductance set $S$. We then run {\Simple}$(G_S,\epsilon_2,\mu_2)$.
Here, the work to output ratio is at most $\otilde(\Delta m^{\tau+\mu_2\tau})$.
If we get a tripartition, the work to output ratio
is at most $\otilde(\Delta m^{1+\mu_1-\tau})$. Adding these, we get an
upper bound on the total work to output ratio.

Because we choose a random subset $P$ of size $O(\log n)$,
we will assume that Lemma~\ref{lem:simple} and Claim~\ref{clm:simple}
hold (without any error).
To analyze the approximation ratio, we follow the progress of the algorithm
to the end. In each iteration, either a low conductance set is removed,
or the basic algorithm is run. In each iteration, let us consider
the set of vertices this is assigned to some side of the final cut.
In case of a low conductance set, we get a cut for the whole set.
Otherwise, if we get a tripartition, the union $Even_i \cup Odd_i$
will be this set. If we do not get a tripartition,
then we output the trivial cut (thereby classifying all remaining vertices).
Let us number the low conductance sets
as $S_1, S_2, \cdots$. The others are denoted $T_1,T_2,\cdots,T_f$.
We will partition the edges of $G$ into parts, defining subgraphs.
The subgraph $G_S$ consists of all edges incident to some
$S_i$. The remaining edges form $G_T$. The
edges of $G_S$ are further partitioned into two sets:
$G_c$ is the subgraph of \emph{cross} edges, which have only
one endpoint in $S$. The other edges make the subgraph $G'_S$.
The edge sets of these subgraphs are $E_S, E_T, E_c, E'_S$,
respectively. For any set $S_i$, $G|_{S_i}$ denotes
the induced subgraph on $S_i$.

We now count the number of edges
in each set that our algorithm cuts. We can only
guarantee that half the edges in $E_c$ are cut. Let the
\maxcut of $G|_{S_i}$ be $\maxcut(G|_{S_i})$. Our algorithm will
cut (in each $S_i$) at least $h(\mu_2)\maxcut(G|_{S_i})$ edges.
This deals with all the edges in $E_S$. In $E_{T_f}$, we can
only cut half of the edges. In $E_{T_j}$, we cut an $H(\epsilon_1,\mu_1)$
fraction of edges. In total,
$$ \sum_i h(\mu_2)\maxcut(S_i) + (1/2)|E_c|
+ \sum_j H(\epsilon_1,\mu_1)|E_{T_j}| + (1/2)|E_{T_f}|$$
The maxcut of $G|_{T_F}$ is at most $(1-\epsilon_1)$ (otherwise, we would
get a tripartition). So we get,
\begin{eqnarray*}
\sum_j H(\epsilon_1,\mu_1)|E_{T_j}| + (1/2)|E_{T_f}| & \geq &
\sum_j H(\epsilon_1,\mu_1)|E_{T_j}| + \frac{1}{2(1-\epsilon_1)} \maxcut(G|_{T_f}) |E_{T_f}| \\
& \geq & \min\big(H(\epsilon_1,\mu_1), \frac{1}{2(1-\epsilon_1)}\big) \maxcut(T)
\end{eqnarray*}

By definition, $|E_c| \leq \chi(\epsilon_1,\mu_1)|E'_S|$. Fixing the size
of $E_c \cup E'_S$, we minimize the number of edges cut by taking
this to be equality. Consider the subgraph $G_S$ and let its
\maxcut value be $1-\epsilon$. If we remove the edges $E_c$,
we get the subgraph $G'_S$. The \maxcut of $G'_S$ is at least
$$ 1 - \frac{\epsilon}{1-\chi(\epsilon_1,\mu_1)} = \frac{1 - \epsilon - \chi(\epsilon_1,\mu_1)}{1-\chi(\epsilon_1,\mu_1)}$$
Now, we lower bound the total number of edges in $G_1$ that are cut.
\begin{eqnarray*} \sum_i h(\mu_2)\maxcut(G|_{S_i}) + (1/2)|E_c| & \geq & h(\mu_2) \sum_i \maxcut(G|_{S_i}) + (1/2)|E_c|\\
& \geq & h(\mu_2) \maxcut(G'_S) + (1/2)\chi(\epsilon_1,\mu_1)|E'_S| \\
& \geq & \big(h(\mu_2) \frac{1-\epsilon-\chi(\epsilon_1,\mu_1)}{1-\chi(\epsilon_1,\mu_1)}
+ (1/2)\chi(\epsilon_1,\mu_1)\big)|E'_S|
\end{eqnarray*}
By definition of $\epsilon$,
$$ \maxcut(G_S) = (1-\epsilon) |E_S| = (1-\epsilon)(1+\chi(\epsilon_1,\mu_1)) |E'_S| $$
The total number of edges cut is bounded below by:
\begin{eqnarray*}
& & \sum_j H(\epsilon_1,\mu_1)|E_{T_j}| + (1/2)|E_f| \\
& \geq & \min_{\epsilon} \left(\frac{1}{2(1-\epsilon)},
\frac{h(\mu_2) (1-\epsilon-\chi(\epsilon_1,\mu_1))}{(1-\epsilon)(1-\chi(\epsilon_1,\mu_1)^2)}
+ \frac{1}{2(1-\epsilon)(1+\chi(\epsilon_1,\mu_1))}\right)\maxcut(G_S) \\
& & + \min\big(H(\epsilon_1,\mu_1), \frac{1}{2(1-\epsilon_1)}\big) \maxcut(G_T)
\end{eqnarray*}
\end{myproof}

Using this we prove the main lemma about {\Balance} (restated here
for convenience):

\begin{Lem}
For any constant $b > 1.5$, there is a choice of $\mu_1$, $\mu_2$ and
$\tau$ so that there is an $\otilde(\Delta m^b)$ time algorithm with an
approximation factor that is a constant greater than $0.5$.
\end{Lem}

\begin{myproof}
The algorithm {\Balance} has a work to output ratio of
$\otilde(\Delta (m^{\tau+\mu_2\tau} + m^{1+\mu_1-\tau}))$. We now set $\mu_1$ and
$\mu_2$ to be constants so that the work to output ratio is $b - 1$. For
this, we set $\tau+\mu_2\tau = 1+\mu_1-\tau = b - 1$. Letting $\mu_1 > 0$
be a free parameter, this gives $\tau = 2 + \mu_1 - b$, and $\mu_2 =
\frac{2b - \mu_1 - 3}{2 + \mu_1 - b}$. Note that since $b >
1.5$, we can choose $\mu_1 > 0$ so that $\tau \geq 0$ and $\mu_2 > 0$.

Now, it remains to show that for any choice of $\mu_1, \mu_2 > 0$, the bound on
the approximation factor given by Claim~\ref{clm:threehalf} is greater than
$0.5$. For convenience of notation, we will drop the arguments to functions and
use $h$, $H$, and $\chi$ to refer to $h(\mu_2)$, $H(\epsilon_1, \mu_1)$, and
$\chi(\epsilon_1, \mu_1)$ respectively. First, note that $h > 0.5$. Let us set
$\epsilon_1 = \bar{\epsilon}(\mu_1)$ as from the statement of
Lemma~\ref{lem:H-properties}. Then $H > 0.5029$, and $\frac{1}{2(1-\epsilon_1)}
> 0.5$ since $\epsilon_1 > 0$. Furthermore, note that
$\min_{\epsilon}\max\left\{\frac{1}{2(1-\epsilon)}, \frac{h
(1-\epsilon-\epsilon\chi) + \chi/2}{(1-\epsilon)(1+\chi)} \right\}$ is obtained
at $\epsilon = \frac{2h - 1}{2h(1 + \chi)}$, and takes the value $\frac{h +
h\chi}{1 + 2h\chi} > 0.5$ since $h > 0.5$. Thus, the minimum of all these three
quantities is greater than $0.5$, and hence the approximation factor is more
than $0.5$.
\end{myproof}\\

Using a more nuanced analysis of the approximation ratio, we can
get better bounds. This requires the solving of an optimization
problem, as opposed to Claim~\ref{clm:threehalf}. We provided
the weaker claim because it is easier to use for Lemma~\ref{lem:balance}.

\begin{claim} \label{clm:lp} Let us fix $\mu_1, \mu_2$. The approximation ratio
can be bounded as follows: let $\epsilon'_S, X, Y, Z$ be variables and $\epsilon, \epsilon_1$ be
fixed. First minimize the function:
$$ \frac{1}{1-\epsilon}\cdot \left[(H(\epsilon'_S,\mu_2) + \chi(\epsilon_1,\mu_1)/2)X
+ H(\epsilon_1,\mu_1) Y + \frac{Z}{2}\right] $$
with constraints:
%
\begin{eqnarray*}
& \epsilon'_S X + \epsilon_1 Z \leq \epsilon \\
& (1+\chi(\epsilon_1,\mu_1))X + Y + Z = 1 \\
& 0 \leq \epsilon'_S \leq 1/2\\
& 0 \leq X, Y, Z \leq 1
\end{eqnarray*}
Let this value by $OBJ(\epsilon,\epsilon_1)$. The approximation ratio is at least
$$\max_{\epsilon_1} \min_{\epsilon} \max[1/(2(1-\epsilon)), OBJ(\epsilon,\epsilon_1)]$$
\end{claim}

\begin{myproof} To analyze the approximation ratio, we follow the progress of the algorithm
to the end. In each iteration, either a low conductance set is removed,
or the basic algorithm is run. In each iteration, let us consider
the set of vertices this is assigned to some side of the final cut.
In case of a low conductance set, we get a cut for the whole set.
Otherwise, if we get a tripartition, the union $V_{i,r}^+ \cup V_{i,r}^-$
will be this set. If we do not get a tripartition,
then we output the trivial cut (thereby classifying all remaining vertices).
Let us number the low conductance sets
as $S_1, S_2, \cdots$. The others are denoted $T_1,T_2,\cdots,T_f$.
We will partition the edges of $G$ into parts, defining subgraphs.
The subgraph $G_S$ consists of all edges incident to some
$S_i$. The remaining edges form $G_T$. The
edges of $G_S$ are further partitioned into two sets:
$G_c$ is the subgraph of \emph{cross} edges, which have only
one endpoint in $S$. The other edges make the subgraph $G'_S$.
In $G_T$, let the edges incident to vertices \emph{not} in
$T_f$ be be $G'_T$. The remaining edges form the subgraph $G_f$.
The edge sets of these subgraphs are $E_S, E_T, E_c, E'_S, E_f, E'_T$,
respectively. For any set $S_i$, $G|_{S_i}$ denotes
the induced subgraph on $S_i$.

We now count the number of edges
in each set that our algorithm cuts. We can only
guarantee that half the edges in $E_c$ are cut. Let the
\maxcut of $G|_{S_i}$ be $\maxcut(G|_{S_i})$ ($= \tau_i$). Our algorithm will
cut (in each $S_i$) at least $H(\tau_i,\mu_2)|E_{S_i}|$ edges.
This deals with all the edges in $E_S$. In $E_{T_f}$, we can
only cut half of the edges. In $E_{T_j}$, we cut an $H(\epsilon_1,\mu_1)$
fraction of edges. In total,
$$ \sum_i H(\tau_i,\mu_2)|E_{S_i}| + (1/2)|E_c|
+ \sum_j H(\epsilon_1,\mu_1)|E_{T_j}| + (1/2)|E_{T_f}|$$
By convexity of $H$, we have
$\sum_i H(\tau_i,\mu_2) \geq H(\epsilon'_S,\mu_2) |E'_S|$,
where $\maxcut(G'_S) = 1 - \epsilon'_S$. Putting it all together, we cut
at least
%
$$ H(\epsilon'_S,\mu_2)|E'_S| + H(\epsilon_1,\mu_1) |E'_T|
+ (1/2)|E_f| +  (1/2)|E_c| $$
We would like to find out the minimum value this can attain, for a given
$\epsilon_1$. The parameters $\mu_1, \mu_2$ are fixed.
The maxcut of $G_f$ is at most $(1-\epsilon_1)$ (otherwise, we would
get a tripartition). We have the
following constraints:
\begin{eqnarray*}
& |E_c| \leq \chi(\epsilon_1,\mu_1)|E'_S| \\
& \epsilon'_S|E'_S| + \epsilon_f|E_f| \leq \epsilon m \\
& |E'_S| + |E'_T| + |E_f| + |E_c| = m \\
& \epsilon_1 \leq \epsilon_f \leq 1/2
\end{eqnarray*}
For a given size of $E'_S$, we should maximize $E_c$ to cut the least
number of edges. So we can assume that $|E_c| = \chi(\epsilon_1,\mu_1)|E'_S|$.
Let us set $X := |E'_S|/m$, $Y := |E'_T|/m$, and $Z := |E_f|/m$.
Consider fixing $\epsilon$ and $\epsilon_1$. The variables are $\epsilon'_S, \epsilon_f, X, Y, Z$.
This means the approximation ratio is at least
the \emph{minimum} of
$$ \frac{1}{1-\epsilon}\cdot \left[(H(\epsilon'_S,\mu_2) + \chi(\epsilon_1,\mu_1)/2)X
+ H(\epsilon_1,\mu_1) Y + \frac{Z}{2}\right] $$
under the constraints:
\begin{eqnarray*}
& \epsilon'_S X + \epsilon_f Z \leq \epsilon \\
& (1+\chi(\epsilon_1,\mu_1))X + Y + Z = 1 \\
& \epsilon_1 \leq \epsilon_f \leq 1/2 \ \ \ 0 \leq \epsilon'_S \leq 1/2\\
& 0 \leq X, Y, Z \leq 1
\end{eqnarray*}
Let $OBJ(\epsilon,\epsilon_1)$ be the minimum value attained.
We observe that given any solution, the objective can
be decreased if we decrease $\epsilon_f$. This is because for
a small decrease in $\epsilon_f$, we can increase $Z$ (and decrease either $X$ or $Y$).
This preserves all the constraints, but decreases the objective.
So we can set $\epsilon_f = \epsilon_1$.
Our bound on the approximation ratio is
$$ \max_{\epsilon_1} \min_{\epsilon} \max[1/(2(1-\epsilon)), OBJ(\epsilon,\epsilon_1)] $$
\end{myproof}

\subsection{Running Time/Approximation Ratio Tradeoff}

\section{Conclusions and Further Work}

Our combinatorial algorithm is very natural and simple, and beats
the $0.5$ barrier for \maxcut. The current bounds for the approximation
ratio we get for, say, quadratic time are quite far from the
optimal Goemans-Williamson $0.878$, or even from Soto's $0.6142$ bound
for Trevisan's algorithm.
The approximation ratio of our algorithm can probably be improved, and it might
be possible to get a better running time. This would probably
require newer analyses of Trevisan's algorithm, similar in spirit
to Soto's work~\cite{Sot09}.
It would be interesting to
see if some other techniques different from random walks
can be used for \maxcut.

This algorithm naturally suggests whether a similar approach can
be used for other $2$-CSPs. We believe that this should be possible,
and it would provide a nice framework for combinatorial algorithms
for such CSPs. On a different note, our local partitioning algorithm
raises very interesting questions. Can we get such a partitioning
procedure that has a better work to output ratio (close to polylogarithmic)
but does not lose the $\sqrt{\log n}$ factor in the conductance
(which previous algorithms lose)? We currently have a work to output
that can be made close to $\sqrt{n}$ in the worst case.
A significant improvement would be of great interest.

\bibliographystyle{alpha}
\bibliography{maxcut}

\end{document}